\documentclass[acmtog]{acmart}

\AtBeginDocument{%
\providecommand\BibTeX{{%
	\normalfont B\kern-0.5em{\scshape i\kern-0.25em b}\kern-0.8em\TeX}}}

\setcopyright{acmlicensed}
\acmJournal{TOG}
\acmYear{2021} \acmVolume{1} \acmNumber{1} \acmArticle{1} \acmMonth{1} \acmPrice{15.00}\acmDOI{10.1145/3480168}

\acmJournal{TOG}
\acmVolume{40}
\acmNumber{4}
\acmArticle{111}
\acmMonth{8}

\usepackage{booktabs} 
\usepackage{gensymb} 
\usepackage{xcolor} 

\usepackage{amsmath}
\usepackage{amsfonts}
\usepackage{subfigure}
\usepackage{array}
\usepackage{booktabs}
\usepackage{multirow}
\usepackage{enumerate}

\usepackage{caption}
\usepackage{graphicx}

\usepackage{hyperref}

\usepackage[normalem]{ulem}

\definecolor{Purple}{cmyk}{0.45,0.86,0,0}
\definecolor{Rosolic}{cmyk}{0.00,1.00,0.50,0}
\definecolor{Brown}{rgb}{0.55,0.27,0.1}
\definecolor{Green}{rgb}{0.1, 0.9, 0.1}

\newcommand{\figref}[1]{Fig.~\ref{#1}}

\newcommand{\equref}[1]{Eqn.~(\ref{#1})}
\newcommand{\secref}[1]{Sec.~\ref{#1}}

\newcommand{\sysname}{GCN-Denoiser}

\begin{document}
	
	\title{\sysname: Mesh Denoising with Graph Convolutional Networks}
	
	\author{Yuefan Shen}
	\email{jhonve@zju.edu.cn}
	\affiliation{%
		\institution{The State Key Lab of CAD\&CG, Zhejiang University}
	}
	\email{jhonve@zju.edu.cn}
	
	\author{Hongbo Fu}
	\affiliation{
		\institution{The School of Creative Media, City University of Hong Kong}
	}
	\email{hongbofu@cityu.edu.hk}
	
	\author{Zhongshuo Du}
	\author{Xiang Chen}
	\affiliation{
		\institution{The State Key Lab of CAD\&CG, Zhejiang University}
	}
	\email{3150104929@zju.edu.cn}
	\email{xchen.cs@gmail.com}
	
	\author{Evgeny Burnaev}
	\affiliation{
		\institution{Skolkovo Institute of Science and Technology}
	}
	\email{e.burnaev@skoltech.ru}
	
	\author{Denis Zorin}
	\affiliation{
		\institution{New York University}
	}
	\email{dzorin@cs.nyu.edu}
	
	\author{Kun Zhou}
	\affiliation{
		\institution{The State Key Lab of CAD\&CG, Zhejiang University}
	}
	\email{kunzhou@acm.org}
	
	\author{Youyi Zheng}\authornote{corresponding author}
	\affiliation{
		\institution{The State Key Lab of CAD\&CG, Zhejiang University}
	}
	\email{youyizheng@zju.edu.cn}

	\renewcommand{\shortauthors}{Shen et al.}
	
	\begin{abstract}
		In this paper, we present \sysname, a novel feature-preserving mesh denoising method based on graph convolutional networks (GCNs). Unlike previous learning-based mesh denoising methods that exploit hand-crafted or voxel-based representations for feature learning, our method explores the structure of a triangular mesh itself and introduces a graph representation followed by graph convolution operations in the dual space of triangles. We show such a graph representation naturally captures the geometry features while being lightweight for both training and inference. To facilitate effective feature learning, our network exploits both static and dynamic edge convolutions, which allow us to learn information from both the explicit mesh structure and potential implicit relations among unconnected neighbors. To better approximate an unknown noise function, we introduce a cascaded optimization paradigm to progressively regress the noise-free facet normals with multiple GCNs. \sysname~achieves the new state-of-the-art results in multiple noise datasets, including CAD models often containing sharp features and raw scan models with real noise captured from different devices. We also create a new dataset called PrintData containing 20 real scans with their corresponding ground-truth meshes for the research community. Our code and data are available in \url{https://github.com/Jhonve/GCN-Denoiser}.
	\end{abstract}

	\ccsdesc[500]{Computing methodologies~Shape analysis}
	\ccsdesc[500]{Computing methodologies~Mesh models}
	
	\keywords{Mesh denoising, graph convolutional networks, cascaded optimization.}
	
	\begin{teaserfigure}
		\includegraphics[width=\textwidth]{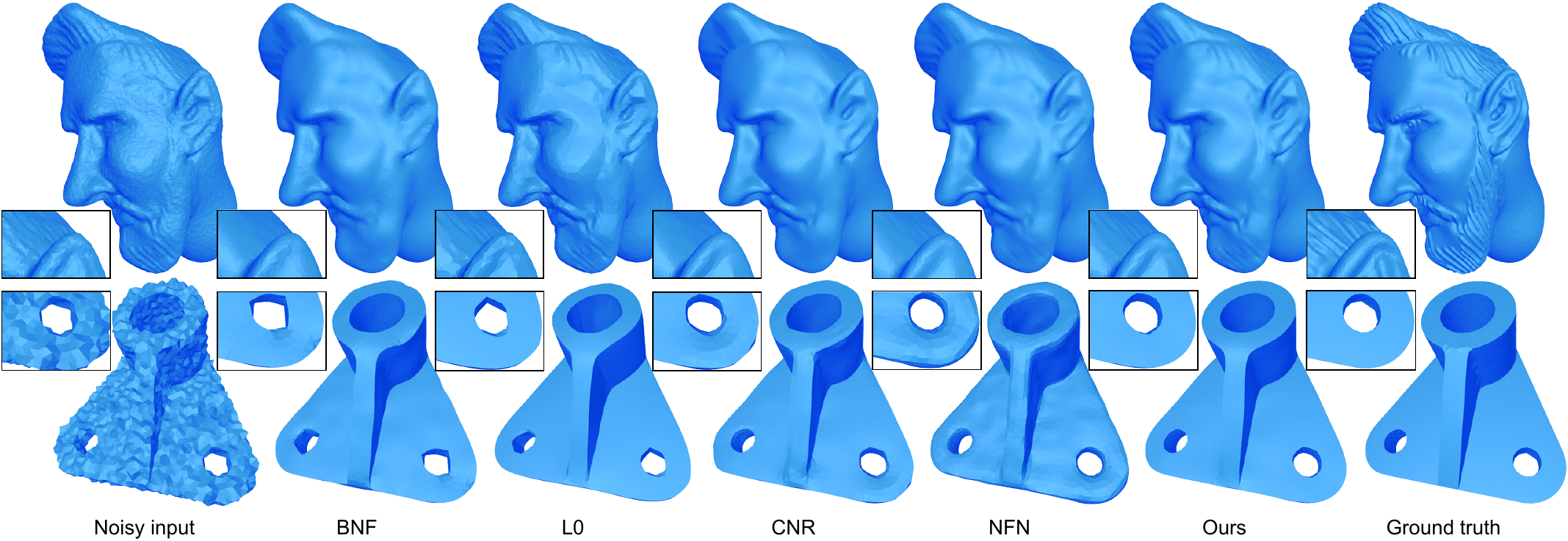}
		\caption{Comparisons of our method to the state-of-the-art mesh denoising methods: Bilateral normal filtering (BNF) \cite{zheng2010bilateral}, $L0$ Smoothing (L0) \cite{he2013mesh},  Cascaded normal regression (CNR) \cite{wang2016mesh}, and NormalF-Net (NFN) \cite{li2020normalf}. Our method consistently achieves the best results for both smooth features and sharp features. The first row shows the denoising results on a real scan model and the second row shows the results on a model with synthetic noise (i.e., Gaussian noise with the level of 0.3 mean edge length). The average normal angular errors {(from left to right)} are: (1$^{st}$ row) 9.68$^{\circ}$, 9.03$^{\circ}$, 8.17$^{\circ}$, 8.29$^{\circ}$, 8.35$^{\circ}$, and \textbf{7.62$^{\circ}$}; (2$^{nd}$ row) 25.85$^{\circ}$, 3.53$^{\circ}$, 5.69$^{\circ}$, 3.36$^{\circ}$, 7.79$^{\circ}$, and \textbf{1.84$^{\circ}$}.}
		\label{fig:teaser}
	\end{teaserfigure}
	
	\maketitle
	
	\section{Introduction}\label{sec:introduction}
	
	With the increasing popularity of consumer-level depth cameras and 3D scanners, it has become easier to acquire 3D models by 3D-scanning real-world objects. However, even with advanced camera technologies, scanned models are inevitably corrupted by noise mainly due to imperfect measurements, making them not immediately usable in the subsequent graphics pipelines. 
	
	Feature-preserving mesh denoising aims to recover the underlying surface signal (e.g., position, normal)  from a measurement with noise $\Delta^*$, essentially, to remove the noise $\epsilon = \Delta^*-\Delta$ while keeping the underlying features preserved. This problem is inherently ill-posed since both $\Delta$ and $\epsilon$ are unknown. To make it tractable, priors and assumptions on $\Delta$ or $\epsilon$ are often made. For example, the pattern of $\epsilon$ is assumed to follow a Gaussian distribution or to be independent and identically distributed. Unfortunately, these assumptions are often invalid for real scanners \cite{wang2012cascaded}, thus challenging a variety of traditional filter-based and optimization-based denoising approaches \cite{fleishman2003bilateral,zheng2010bilateral,zhang2015guided,he2013mesh,li2018non,wei2018mesh}.
	
	Data-driven methods have attracted a lot of attention lately and several approaches have been introduced for mesh denoising \cite{wang2016mesh,zhao2019normalnet,li2020normalf}. Without making any specific assumption on the data, these methods estimate $\epsilon$ from massive noisy meshes and their ground-truth counterparts, and achieve impressive results. However, their learning paradigms still suffer from several aspects. First, since $\epsilon$ can be a highly complex function eroded over various geometric features, globally estimating $\epsilon$ is often intractable and thus surface patches are usually exploited to regress $\Delta$ locally. This makes the design of a representation of local patches an essential issue for learning-based algorithms. Existing data-driven methods either use hand-crafted features \cite{wang2016mesh,li2020normalf} or exploit a voxel representation \cite{zhao2019normalnet}, leading to either insufficient or redundant information flow into the subsequent learning module. In addition, methods that rely on convolution operations on voxels are known to be slow \cite{OCNN2017}, causing an additional efficiency issue.
	
	To seek the balance between efficacy (geometry representation) and efficiency, we present \emph{\sysname}, a novel approach for feature-preserving denoising of mesh surfaces (triangular meshes in our implementation) based on the extension of graph convolutional networks (GCNs), which have been proved effective for various applications \cite{schult2020dualconvmesh} but not for feature-preserving mesh denoising. \sysname~exploits a novel graph-based representation for local surface patches and employs graph convolution operations to effectively learn the relations between $\epsilon$ and $\Delta$. The graph representation naturally fits into the local structure surrounding a surface facet and thus has better capability in capturing the local geometric information than hand-crafted features and voxel-based features, enabling a more accurate estimation of $\epsilon$ while being also efficient at runtime.
	
	Like previous works \cite{he2013mesh,wang2016mesh}, we model $\Delta$ as surface normal signals. Our goal is to regress for each facet a normal from a local patch surrounding the facet. To this end, we build our graph structure in the dual space of mesh facets, where each facet of the mesh forms a graph node and adjacent facets are connected with graph edges. We then define graph convolution operations that directly operate on mesh facets to deal with varying local graph structures. To effectively learn features, our network architecture employs both static and dynamic edge convolution operations to exploit both the explicit and implicit graph structures, enabling information flow from both the neighboring facets and the unconnected ones in a patch. We further introduce a normal tensor voting strategy to make our patch representation rotation-invariant. Finally, cascaded optimization is employed to progressively regress $\Delta$ using multiple GCNs.

	We test our method extensively on both synthetic and real-scan models (including a newly created dataset: PrintData) and make comparisons with the state-of-the-art algorithms. Various experiments demonstrate that our method produces superior results to the compared approaches. In summary, the main contributions of this paper are:
	\begin{itemize}
		\item We introduce the first GCN-based method for feature-preserving mesh denoising, which achieves the new state-of-the-art results while being well-balanced between efficacy and efficiency;
		\item We present a rotation-invariant graph representation on the dual space of mesh faces, enabling effective feature learning with graph convolutions; 
		\item We create a new feature-preserving mesh denoising dataset consisting of 20 real scans	with corresponding ground-truth meshes by using a high-end 3D printer and a high-resolution scanner.
	\end{itemize}
	
	\begin{figure*}[t]
		\centering
		\includegraphics[width=\linewidth]{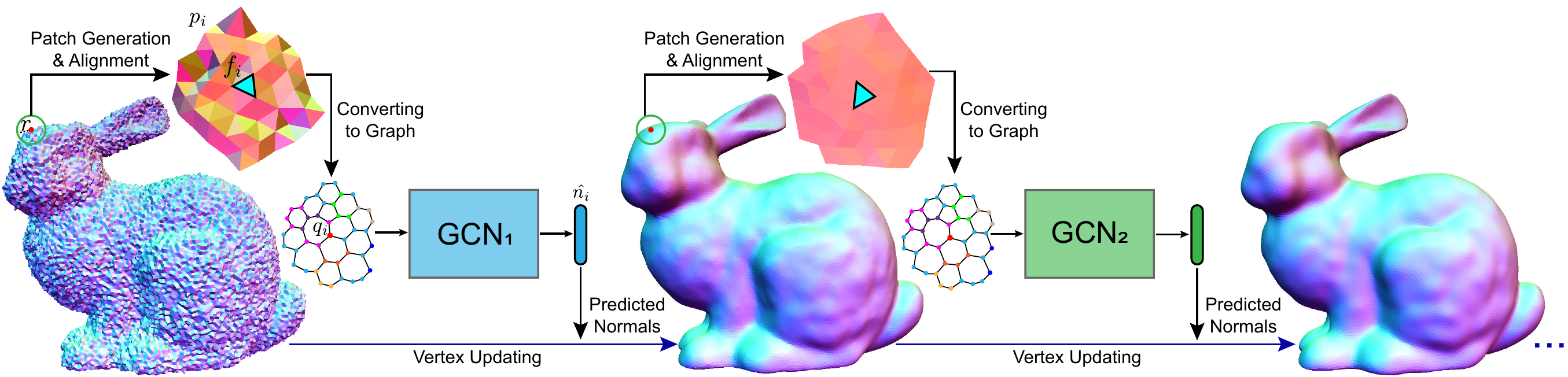}
		\caption{The pipeline of \sysname. For each mesh facet, a graph patch is generated in the dual space of triangles and subsequently fed into multiple GCNs to regress the final noise-free normal for the facet. The output normals of each GCN are used to update the mesh geometry via vertex updating. Then each updated graph patch is fed into the next GCN. In practice, 2 GCNs are sufficient to achieve sufficiently denoised results.}
		\label{fig:framework}
	\end{figure*}
	
	\section{Related Work}
	We review the literature of feature-preserving mesh denoising and some recent advances in graph convolutional networks.
	
	\subsection{Mesh Denoising}
	The task of denoising on mesh surfaces is akin to that on 2D images, since vertex positions or face normals are essentially signals in 3D. Hence, mesh denoising techniques have been heavily inspired by denoising techniques in images. Various low-pass and feature-preserving filters have been introduced for mesh denoising \cite{clarenz2000anisotropic, tasdizen2002geometric, bajaj2003anisotropic, yagou2003mesh,yagou2002mesh, shen2004fuzzy}, based on the assumption that the noise over a surface is often of high frequency. Among them, a bilateral filter is one of the most widely applied filters \cite{jones2003non, fleishman2003bilateral,shimizu2005new,Shin2006,fan2009robust,adams2009gaussian,zheng2010bilateral,wang2012cascaded,wei2014bi,yadav2018robust} due to its simplicity and feature-preserving property. It is reported in \cite{lee2005feature,zheng2010bilateral} that normals are often more discriminative in describing local geometry than vertex positions. In light of this, a series of research works improve the denoising performance by first updating facet normals and then vertex positions \cite{sun2007fast}. Extensions to these works have been made later on to find more robust filtering schemes on facet normals \cite{zhang2015guided, zhao2019graph}. A common drawback in filter-based methods is that once the features are highly corrupted by noise, they, especially weak ones, are difficult to {be recovered} by these methods.
	
	Optimization-based mesh denoising methods take from another perspective by seeking a denoised mesh to approximate an input mesh conditional on a set of priors imposed on the ground-truth geometry or noise patterns. They formulate the denoising procedure as optimization problems and solve them via techniques such as Bayesian \cite{Bayesian2006}, $L_0$ minimization \cite{he2013mesh}, compressed sensing \cite{wang2014decoupling}, or low-rank recovery \cite{li2018non,wei2018mesh}. These methods, however, only work well for meshes where their assumptions are satisfied, and often do not generalize well for meshes with different geometry features and noise patterns.
	
	In contrast, learning-based methods do not make specific assumptions about the underlying features or noise patterns,  and have been successfully applied to image denoising \cite{burger2012image, xie2012image, agostinelli2013adaptive, zhang2017beyond, jin2017deep,yan2018ddrnet,sterzentsenko2019self}. However, unlike images, 3D meshes are usually irregular, thus making image-based convolutional operations not directly applicable. To address this issue, Wang et al. \shortcite{wang2016mesh} introduce a filtered facet normal descriptor (FND), which is extracted from a local region by bilateral filtering with multiple kernels, and utilize simple multilayer perceptrons (MLPs) to regress noise-free normals from the descriptor. Later on, Zhao et al. \shortcite{zhao2019normalnet} employ a voxel-based representation and apply 3D convolution to regress noise-free normals. In \shortcite{li2020normalf}, Li et al. adopt a non-local patch matrix proposed by \cite{li2018non} as a regular representation of mesh patches and use 2D convolutional networks to learn noise-free normals. Wei et al. \shortcite{wei2019mesh} introduce a learning-based de-filtering strategy to recovery over-smoothed weak features, but their results depend on the quality of pre-denoised meshes. In a concurrent work, Li et al. \shortcite{li2020dnf} omit the adjacent relationship between mesh faces and use a similar architecture of PointNet++ \cite{qi2017pointnetplus} to regress a new normal vector from a patch of facet normals. Different from these works, our method directly feeds an irregular mesh patch data into a graph convolutional network rather than seeking an intermediate hand-crafted representation. We show that our graph-based representation of local surface geometry with graph convolution operations enables our network to have better capability in capturing the inherent geometry features of a noisy patch than the above-discussed methods.
	
	\subsection{Graph Convolutional Networks}
	Graph convolutional networks (GCNs) have been introduced for handling non-Euclidean structures. Early works of GCNs require a static graph structure \cite{bruna2013spectral, defferrard2016convolutional} and thus cannot be extended for meshes with varying topology. Recent studies on dynamic graph convolution show that changeable edges can perform better. For example, Simonovsky et al. \shortcite{simonovsky2017dynamic} generalize the convolution operator to irregular graphs with filter weights. Wang et al. \shortcite{wang2019dynamic} dynamically construct graph structures by connecting neighboring nodes in each layer. Valsesia et al. \shortcite{Valsesia:2019:LLGM} construct node neighbors using K-nearest-neighbors (KNN). Furthermore, Li et al. \shortcite{li2019deepgcns} show that very deep GCNs with dynamic graph convolutions can further boost the performance in applications such as point cloud recognition and segmentation. Our method exploits both the static graph structure in patches and dynamic graph structures that are constructed during convolution to effectively learn the geometry features in patches.
	
	There are some other convolution operators developed for meshes lately. In \cite{masci2015geodesic, monti2017geometric}, local coordinate systems are defined to confine the convolution operations on regular grids. MeshCNN \cite{hanocka2019meshcnn} executes convolution or pooling on mesh edges while \cite{schult2020dualconvmesh} performs graph convolution and vertex pooling on mesh vertices. In \shortcite{feng2019meshnet}, Feng et al. introduce a convolution on mesh facets and separate mesh features into spatial and structure levels manually. These methods are mainly designed for understanding whole objects or large scenes and require very deep architectures. Instead, we pay more attention to local patches and employ the convolution in the dual space of mesh facets.
	
	\section{Algorithm Overview}\label{sec:overview}
	Fig. \ref{fig:framework} illustrates the pipeline of \sysname. Since $\epsilon$ is usually a highly complex function over the underlying surface, we take a local approach to approximate it. Our aim is to predict a noise-free normal $n_f$ for each individual facet $f$ surrounded by a noisy local surface patch $p$ in the dual domain of mesh triangles. Given the denoised facet normals, we update the vertex positions according to the method of \cite{zheng2010bilateral} to get a denoised surface. 
	
	Given a noisy triangular mesh, we first generate a local patch for each facet. To eliminate the global spatial transformations among the local patches, we apply normal tensor voting to the collected patches to align them into a common embedding (Section \ref{subsec:tv}). Next, we convert these aligned patches into graph representations (Section \ref{subsec:graph}), then feed these patch graphs into our GCN network to progressively compute the noise-reduced normals (Section \ref{subsec:gcn}), {and then obtain a noise-reduced mesh via vertex updating (Section \ref{sec:denoise})}. Even when we approximate the noise function $\epsilon$ in local patches, the underlying noise patterns there could still be complex as they are eroded by different geometric features. Hence, as a common practice for approximating a highly nonlinear function \cite{wang2016mesh,li2020normalf}, we employ a cascaded optimization to train multiple GCNs to progressively regress the final noise-free normals. In each GCN, pairs of $(p,n_f)$ are collected for training (Section \ref{subsec:data}). The overall workflow is similar among all the cascaded stages, and the only difference is that the noise levels of $\{p\}$-s are different.
	
	\section{Patch Generation and Alignment}\label{sec:patch}
	We define an input triangular mesh as $M=\{V, F\}$, with $V=\{v_i\}_1^{N_v}$ the set all vertices and $F=\{f_i\}_1^{N_f}$ the set all facets. $N_v$ and $N_f$ are respectively the number of vertices and the number of faces. For each facet $f_i$ in $F$, we will generate its local patch data $p_i$. The set of all patches in $M$ is defined as $P=\{p_i\}_1^{N_f}$. Also, we denote the normal of a facet $f_i$ as $n_i$, its centroid as $c_i$, and its area as $a_i$.
	
	\subsection{Patch Selection}\label{subsec:select}
	A patch $p_i$ in our context refers to all facets (including $f_i$) within a sphere of radius $r$ located at the {centroid} $c_i$ of facet $f_i$, i.e., $p_i$ is supposed to satisfy:
	\begin{equation}\label{eqn:patchsize}
		\mathop{\forall}\limits_{f_j\in{p_i}}\mathop{\exists}\limits_{v_k\in{f_j}}||v_k - c_i|| < r.
	\end{equation}
	We use $r = k\bar{a}_i^\frac{1}{2}$, where $\bar{a}_i$ is the average area of 2-ring neighboring facets of $f_i$ and $k$ is a parameter relevant to the resolution of an input mesh (Section \ref{subsec:ablation}). We incorporate the facet area in defining our patch to accommodate irregular sampling (see an example in Fig. \ref{fig:normaliter}). 
	
	\subsection{Patch Alignment via Tensor Voting}\label{subsec:tv}
	Patches at different spatial locations would cause troubles for neural networks, since learning spatial transformations is known to be difficult for deep methods {\cite{rotation_invariant2020}}. To address this issue, we resort to a normal tensor voting strategy to explicitly align the patches into a common coordinate system.
	
\begin{figure}[t!]
		\centering
		\includegraphics[width=\linewidth]{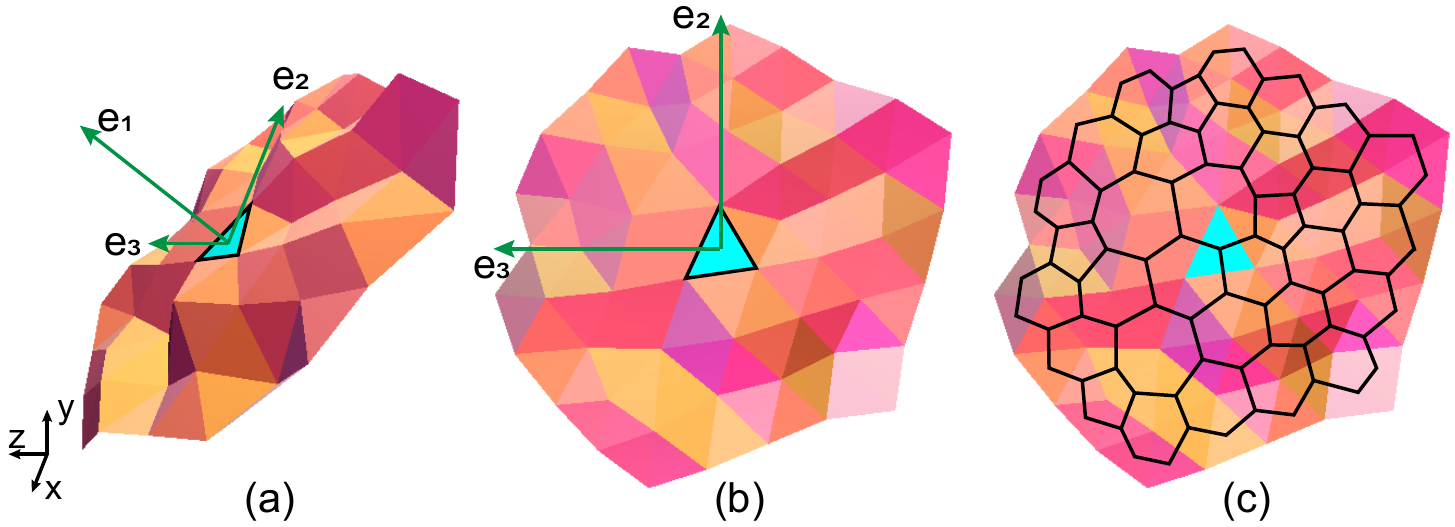}
		\caption{(a) original patch, (b) patch after applying our normal tensor voting to (a), and (c) the graph structure of the patch.}
		\label{fig:align}
	\end{figure}

	As in \cite{zhao2019normalnet}, we first translate $p_i$ to the origin $[0, 0, 0]$ and also scale it into a unit bounding box. The normal voting tensor $T_i$ \cite{shimizu2005new} for facet $f_i$ is then defined by:
	\begin{equation}
		\label{eq:tv}
		T_i = \displaystyle\sum_{f_j \in {p_i}} \mu_j{n_j}'{{n_j}'}^T,
	\end{equation}
	where $\mu_j = (a_j/a_m)\exp{(-||c_j-c_i||/\sigma)}$, with $a_m$ being the maximum triangle area in $p_i$, and ${n_j}'$ is the voted normal of $f_j$: ${n_j}'=2(n_j\cdot{w_j})w_j-n_j$, with $w_j=normalize\{[(c_j-c_i)\times{n_j}]\times{(c_j-c_i)}\}$. Since $T_i$ is a positive semi-definite matrix and can be represented by its spectral decomposition as:
	\begin{equation}
		\label{eq:ti}
		T_i = \lambda_1e_1{e_1}^T+\lambda_2e_2{e_2}^T+\lambda_3e_3{e_3}^T,
	\end{equation}
	where $\lambda_1\geq\lambda_2\geq\lambda_3$ are its eigen values, and $e_1, e_2$ and $e_3$ are the corresponding unit eigen vectors, which form a group of orthonormal basis. Then we construct a rotation matrix $R_i=[e_1, e_2, e_3]$ and multiply each facet's centroid and normal in $p_i$ with $R_i^{-1}$ to generate new patch data $\bar{p_i}$. We show the comparison between the rendered patch data before and after alignment in Fig. \ref{fig:align}. 
	
	\subsection{Graph Representation}\label{subsec:graph}
	
	We build a graph structure for each of our generated patches after alignment in order to shape it to fit our subsequent graph convolutional networks. An undirected graph $\mathcal{G}=(Q, E, \Phi)$ is built, where a graph node $q_i \in Q$ is created for each facet $f_i$ in patch $\bar{p}$ and an edge $e = (q_i,q_j) \in E$ is created if the corresponding faces $f_i$ and $f_j$ are adjacent. Fig. \ref{fig:align} (c) shows an example. $\Phi$ contains a set of node attributes, i.e., feature tuples. For each $\phi_i \in \Phi$, which corresponds to facet $f_i$, we set $\phi_i = (\bar{c}_i,\bar{n}_i,a_i,d_i)$. Here $\bar{c}_i$ and $\bar{n}_i$ respectively indicate the centroid and normal of facet $f_i$ after alignment. $d_i$ is the number of neighboring facets in the 1-ring neighborhood of $f_i$, which helps distinguish boundary faces.
	
	\section{Normal Regression}\label{sec:learn}
	With the graph representation $\mathcal{G}_i$ extracted for each patch $p_i$, we now detail our multiple graph convolutional networks that take $\mathcal{G}_i$ as input and output a denoised normal vector $\hat{n_i}$ for facet $f_i$.
	
	\begin{figure}[b!]
		\centering
		\includegraphics[width=\linewidth]{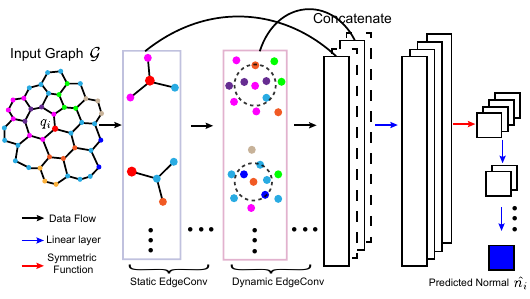}
		\caption{Our GCN architecture. 
			Both static EgdeConv and dynamic EdgeConv are applied and their aggregated information is combined to flow into subsequent MLPs to regress a noise-reduced normal.
		}
		\label{fig:gcn}
	\end{figure}
	
	\subsection{Graph Convolutional Network}\label{subsec:gcn}
	Our GCN network consists of multiple convolutional layers. In each layer, similar to traditional convolutional networks, our GCN aggregates features from neighboring nodes of each node and updates its features. The aggregating and updating are also named as a convolutional operation.
	
	\paragraph{Graph Convolution}
	Although each graph contains necessary connectivity among facets, their structures vary across patches. Thus, we adopt an Edge-Conditioned Convolution (ECC) strategy proposed in \cite{simonovsky2017dynamic} to deal with convolutions among varying structures. Specifically, we use \emph{EdgeConv} \cite{wang2019dynamic} as described below.
	
	Let {$\mathcal{G}_l=(Q_l, E_l, \Phi_l)$} be the $l$-th layer in our graph convolutional network and $\mathrm{F}_l^i$ is the feature vector of the $i$-th node in $\mathcal{G}_l$. EdgeConv updates nodes features by:
	\begin{equation}
		\label{eq:edgeconv}
		\mathrm{F}_{l+1}^i = \mathop{\Psi}\limits_{j:(i,j)\in{E_l}}h_\Theta^{l}(\mathrm{F}_l^{i}, \mathrm{F}_l^{j}), 
	\end{equation}
	where $\Psi$ is a max aggregation operation and $h_\Theta^{l}=\mathtt{Linear}_\Theta^l(\mathrm{F}_l^{i}, \mathrm{F}_l^{j}-\mathrm{F}_l^{i})$. Each graph convolutional layer of our network shares the same $\mathtt{Linear}_\Theta$, which is a multi-layer perceptron (MLP) including batch normalization and activation function $\mathtt{LeakyReLU}$.
	
	By only utilizing the original graph structure might lead to some missing information during convolution since the mapping from geometry to connectivity is not a one-to-one function. To enrich the receptive field of a graph node, we further allow non-adjacent graph nodes to be connected during convolution. This corresponds to a dynamic graph construction procedure \cite{simonovsky2017dynamic,wang2019dynamic}. We call graph convolution with this scheme \emph{dynamic EdgeConv}. For this scheme, the neighbors of each node are dynamically calculated by KNN (K = 8 in our implementation) based on the {Euclidean} distance of node features.
	
	\paragraph{Network Architecture and Training}
	As shown in Fig. \ref{fig:gcn}, our network architecture consists of $L_e$ layers of EdgeConv, $L_d$ layers of dynamic EdgeConv, and $L_l$ layers of fully connected (FC) layers. After the layers of graph convolution, the learned features are concatenated together for pooling. We use both average pooling and max pooling as symmetric functions, which are able to select the most important features. Finally, the FC layers follow to regress a 3D vector, which is our predicted normal. Every layer in our architecture except the last FC layer is followed by batch normalization and activation function $\mathtt{LeakyReLU}$.
	
	As discussed in Section \ref{sec:overview}, we use cascaded GCNs (GCN$_1$, ..., GCN$_l$) to progressively regress the noise-free normals. All GCNs used in the cascaded optimization share the same architecture but with different numbers of EdgeConv, dynamic EdgeConv, and FC layers and are trained iteratively as in \cite{wang2016mesh}. In our experiments, we use $L_e = 3, L_d = 3$, and $L_l = 4$ for the first GCN, and $L_e = 2, L_d = 2$, and $L_l = 3$ for the rest, since we wish the first one to recover the normals coarsely while the rest ones to refine the details. The loss function is MSE between the network output and the ground truth normal, which is $R^{-1}\tilde{n}_f$. Here, $\tilde{n}_f$ is {the} normal of face $f_i$ in the ground-truth noise-free mesh and $R$ is the corresponding rotation matrix in Section \ref{subsec:tv}. We use the MSE loss instead of cosine similarity since it leads to more stable training as MSE imposes hard constraints on the values to be within the range of (0,1). Note that we map the normalized normals from (-1, 1) into (0, 1).
	
	In an offline training step, we use the output of GCN$_i$ to denoise (with both normal updating and vertex updating) a noisy mesh in our training set, and new graphs are generated from these updated meshes to train the next GCN$_{i+1}$. We stop cascading our GCNs when the validation error is not decreasing. In our experiments, we find that from GCN$_2$, the accuracy of GCN$_i$ improves very slightly (see in Fig. \ref{fig:cascade}), thus, unless explicitly indicated, we use two cascaded GCNs in our pipeline.
	
	\begin{figure}[t!]
		\centering
		\includegraphics[width=\linewidth]{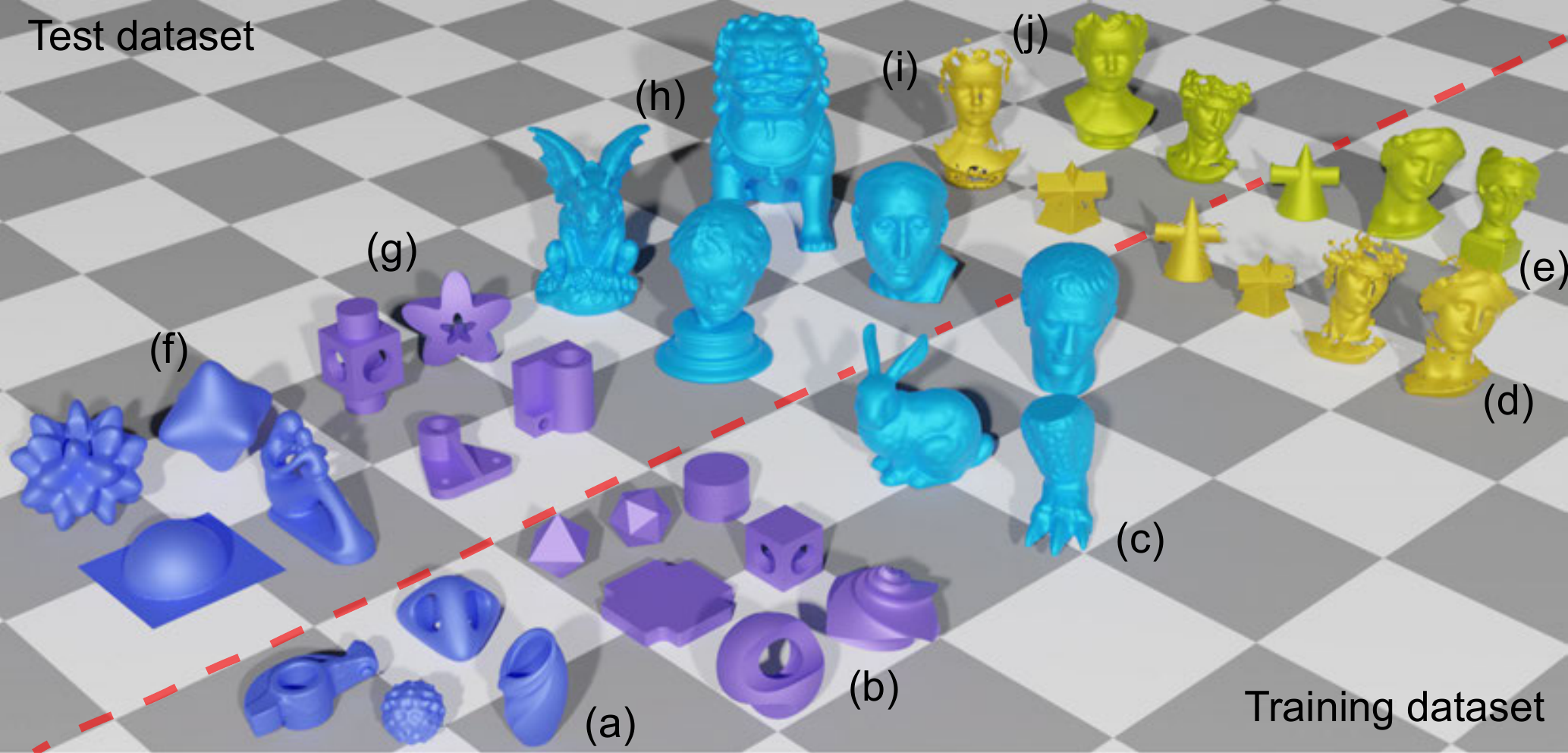}
		\caption{Representative mesh models including (a, b, c, f, g, h) those with synthetic noise (SysData), (d, i) Kinect v1/v2 real scan models (Kv1Data, Kv2Data), and (e, j) Kinect v1 fused models (K-FData). Following \cite{wang2012cascaded}, we consider (a, f) as smooth models, (b, g) CAD models, and (c, h) feature models. The bottom-right set of models consists of all synthetic models and sampled real scan models for training, and the top-left set of models consists of sampled test models.
		}
		\label{fig:dataset}
	\end{figure}
	
	\begin{figure}[b!]
		\centering
		\includegraphics[width=\linewidth]{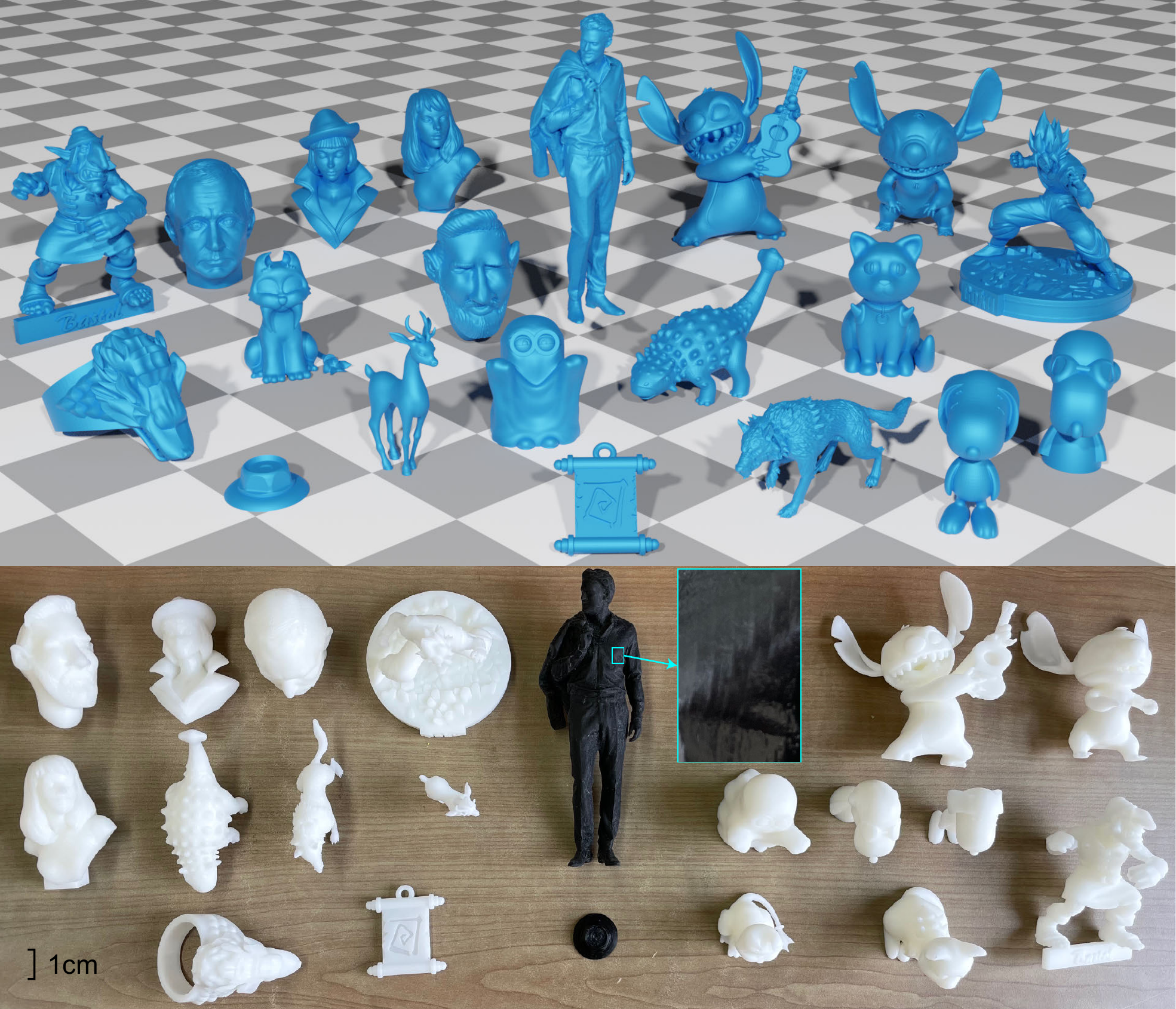}
		\caption{Mesh models and printed results in the proposed PrintData dataset, consisting of 20 models in total. The close up shows some print errors like the vertical lines.
		}
		\label{fig:printdataset}
	\end{figure}
	
	\begin{figure}[b!]
		\centering
		\includegraphics[width=\linewidth]{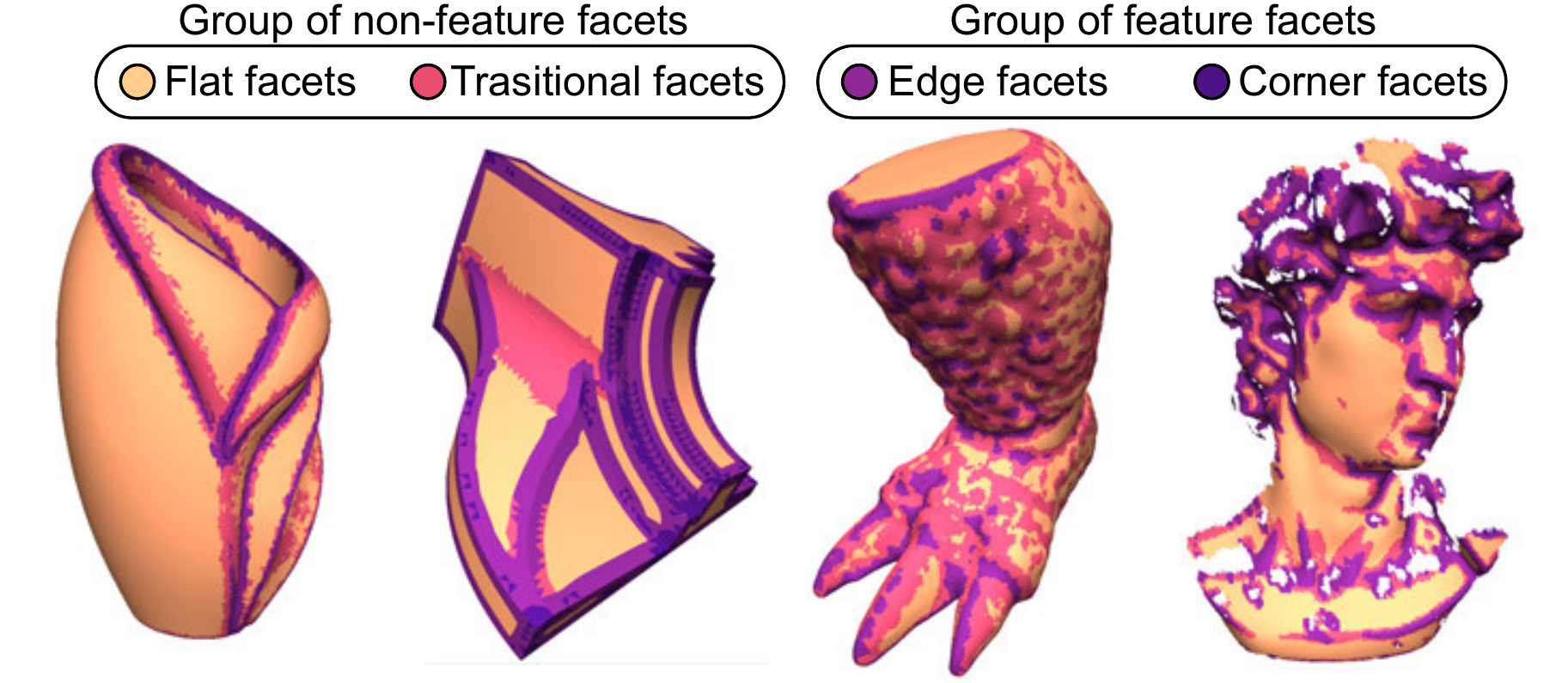}
		\caption{We classify the facets into two main groups: a group of non-feature facets (flat facets and transitional facets) and a group of feature facets (edge facets and corner facets), and select samples from these two groups for a balanced training.}
		\label{fig:classify}
	\end{figure}
	
	\subsection{Data Generation}\label{subsec:data}
	
	Our training set contains mesh models with both synthetic or real noise (Fig. \ref{fig:dataset}). For each 3D model, we generate different levels and types (Gaussian and Impulsive) of noise for training.
	
	We observe that although mesh surfaces appear rather different from each other, their local patches are less diverse. For example, in most of the CAD models, there are a redundant number of flat patches, and edge and corner features, which would cause an imbalanced data distribution during training. In light of this, we apply the tensor voting strategy described in Section \ref{subsec:tv} on each facet of noise-free models in our dataset and get three eigenvalues $\lambda_1, \lambda_2$, and $\lambda_3$. For each model we first classify the facets into four groups: those with $\{f_i|\lambda_2^i < 0.01 \wedge \lambda_3^i < 0.001\}$ as flat facets, those with $\{f_i|\lambda_2^i > 0.01 \wedge \lambda_3^i < 0.1\}$ as edge facets, those with $\{f_i|\lambda_3^i > 0.1\}$ as corner facets, and the rest as transitional facets. Several examples of classified faces are shown in Fig. \ref{fig:classify} with different colors. Due to the numbers of edge and corner facets are smaller compared with those of the other two types of features, we further arrange them into two groups: a group of non-feature facets composed of flat and transitional facets, and a group of feature facets composed of corner and edge facets. We then sample among the two groups to collect our training patches. This data-balancing strategy is used for training all our GCNs. 
	
	\section{Surface Denoising With Predicted Normals}\label{sec:denoise}
	Given the predicted normals of a noisy input mesh, we first refine them to handle possible discontinuities between adjacent faces and then reconstruct its denoised version under their guidance. 
	
	\begin{figure}
		\centering
		\includegraphics[width=\linewidth]{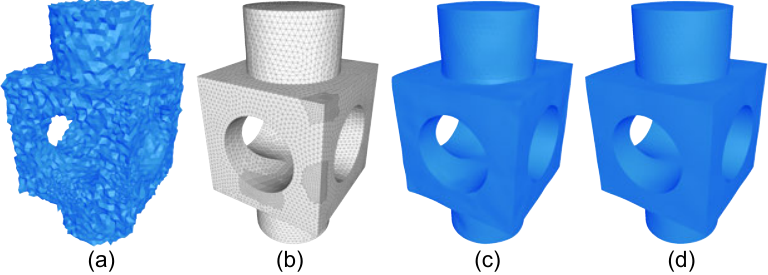}
		\caption{Normal refinement further improves the output normals from our GCNs output (c), and leads to a slightly better denoised result (d). (b) is the wireframe of the original noise-free mesh showing the irregular sampling. The average angular errors $E_a$ of (c) and (d) are 2.67$^{\circ}$ and \textbf{2.16$^{\circ}$}, respectively.}
		\label{fig:normaliter}
	\end{figure}
	
	\subsection{Normal Refinement}\label{subsec:normalupdate}
	The method based on local regression of each face's normal independently would lead to a problem of discontinuities between adjacent faces. This results in small perturbations on the denoised surface especially for some CAD models (Fig. \ref{fig:normaliter}). To alleviate this issue, we apply a bilateral filter \cite{zheng2010bilateral} with small kernels to refine the normals predicted by cascaded GCNs:
	\begin{equation}
		\label{eq:bilateral}
		\dot{n}_i = normalize(\displaystyle\sum_{f_j\in\Omega_i}a_jW_s(||c_j-c_i||)W_r(||\hat{n}_j-\hat{n}_i||)\hat{n}_j),
	\end{equation}
	where $\dot{n}_i$ is the final refined normal, $\Omega_i$ is a set of neighbors of $f_i$ defined in \cite{zhang2015guided}, $W_s$ and $W_r$ are Gaussian functions with kernels $\sigma_s$ (spatial variance) and $\sigma_r$ (range variance) respectively, and $\hat{n}$ represents the predicted normal outputted by our GCNs. The refinement is applied iteratively with $m$ steps and fixed kernels ($\sigma_s=\bar{l}_e$, where $\bar{l}_e$ is the average distance between neighboring facet centroids across the whole mesh and $\sigma_r = 0.3$). Note that this normal refinement is only applied to the output normals of the last cascaded GCN. 
	
	In our experiments, we find that meshes  consisting of many large flat areas of features or corrupted by high noise can be refined well after $m \in [8,16]$ iterations of normal refinement, while meshes with fine features and small noise do not necessarily need normal refinement. Thus, in our experiments, we set $m = 12$ for CAD models and Kinect real-scan models, which often contain high noise, and $m = 1$ for the others. Fig. \ref{fig:synt_quantitative} and \ref{fig:real_quantitative} show the effect of normal refinement.
	
	\subsection{Vertex Updating}\label{subsec:vertexupdate}
	The vertex updating scheme in our method is the same as \cite{sun2007fast, zheng2010bilateral}, and defined by:
	\begin{equation}
		\label{eq:vertex}
		v_i^{k+1} = v_i^{k}+\frac{1}{3|\Omega'_i|}\displaystyle\sum_{f_j\in{\Omega'_i}}\sum_{e_{ij}\in\partial{f_j}}n_j^g[n_j^g\cdot(v_j^{k}-v_i^{k})],
	\end{equation}
	where $\Omega'_i$ is a set of one-ring neighboring faces for $v_i$ and $n_j^g$ is the denoised normal of {facet $f_j$}. In our experiments, we find that 15 iterations are sufficient for all our results.
	
	\section{Experiments}\label{sec:exp}
	We have conducted extensive experiments to evaluate our proposed method with both synthetic and real-scan datasets. All of our experimental tests are conducted on a PC with Intel(R) Core(TM) i7-8770 3.20GHz CPU, 16GB memory, and one GeForce GTX 1080Ti GPU. Our GCNs are implemented using the PyTorch framework and integrated with the PyTorch C++ Library.
	
	\subsection{Dataset}\label{subsec:dataset}
	We build our training datasets on the models from \cite{wang2016mesh} (see representative models in Fig. \ref{fig:dataset}), consisting of:
	
	\begin{itemize}
		\item {SysData}: There are 14 models as shown in {(a), (b) and (c)} of Fig. \ref{fig:dataset}. For each mesh, we synthesize three levels of Gaussian (with their deviations set to 0.1, 0.2, and 0.3 of each mesh average edge length) and impulsive noise (the numbers of impulsive vertices are 10\%, 20\%, and 30\% of the mesh vertex numbers) for training. After data balancing (Section \ref{subsec:data}), there are about 2.4M patches in this dataset. This dataset contains three types of mesh models: CAD models, smooth models, and models with rich fine-scale features as shown in Fig. \ref{fig:dataset}.
		\item {Kv1Data}: This set contains meshes scanned by Microsoft Kinect v1. We select 48 frames from four scanned models (12 frames for each model) and there are about 910K patches in total after data balancing.
		\item {Kv2Data}: These are meshes scanned by Microsoft Kinect v2 and we select 48 frames from four scanned models (12 frames for each model). In total, there are about 560K patches after data balancing.
		\item {K-FData}: This set contains meshes scanned by Microsoft Kinect v1 via the Kinect-Fusion technique \cite{KinectFusion2011}. We generate about 200K patches from 3 models (see in Fig. \ref{fig:dataset} e).
	\end{itemize}
	For the real scan datasets (i.e., Kv1Data, Kv2Data, and K-FData), their ground-truth counterparts are provided by \cite{wang2016mesh}.
	
	For testing, four benchmark datasets are paired with the training datasets accordingly. For SysData, similarly, three levels of Gaussian noise are added onto 29 meshes including 14 CAD-like mesh models, 7 smooth mesh models, and 8 mesh models with rich features, so there are 87 models in total. For real-scan datasets, there are 73 scanned frames for the Kv1Data benchmark, 72 scanned frames for the Kv2Data benchmark, and 4 models for the K-FData benchmark. Apart from these, we build a new real-scan dataset, consisting of 20 scans and their corresponding ground-truth models:
	\begin{itemize}
		\item {PrintData}: To create this dataset, we first select 20 3D models from an online 3D model repository (\url{3dmag.org}). These original 3D models serve as the ground truth. We then prepare physical models by 3D-printing these digital models using a high-end 3D printer (Stratasys Eden260v) and scan the printed models using high-resolution scanners (Artec $\text{Spider}^{\text{TM}}$ and SHINNING 3D EinScan Pro 2x).
	\end{itemize}
	
	\subsection{Error Metrics}
	To evaluate our results and quantitatively compare our method with the state-of-the-art methods, we use two commonly adopted types of metric defined as follows. $E_a$ measures the average normal angular difference between a denoised mesh and the original noise-free mesh.
	\begin{equation}\label{eq:ea}
		E_a=\frac{1}{N_f} \displaystyle\sum_{f_i^r\in{F^r}}acos(n_i^r\cdot\tilde{n}_i),
	\end{equation}
	where $n_i^r$ and $\tilde{n}_i$ are a normalized normal of the $i$-th facet in the denoised mesh and the normalized normal of the corresponding facet in the ground-truth mesh. {$F^r$ is the set of all facets in the denoised mesh.} 
	\begin{equation}\label{eq:ev}
		E_v=\frac{1}{N_vL_d} \displaystyle\sum_{v_i^r\in{V_M^r}} \mathop{\min}\limits_{\tilde{v}_j\in\tilde{V}_M}||v_i^r - \tilde{v}_j||. 
	\end{equation}
	Here $E_v$ is the normalized average Hausdorff distance from a denoised mesh to the corresponding ground-truth 
	mesh \cite{wang2016mesh}, where $L_d$ is the diagonal length of the mesh's bounding box, and $V_M^r$ and $\tilde{V}_M$ are vertices sets of these paired meshes after Monte Carlo sampling. 

	\subsection{Results and Comparisons}
	We compare both qualitatively and quantitatively of our method with the state-of-the-art mesh denoising methods including bilateral mesh filtering (BMF) \cite{fleishman2003bilateral}, bilateral normal filtering (BNF) \cite{zheng2010bilateral}, the $L_0$ smoothing (L0) \cite{he2013mesh}, guided normal filtering (GNF) \cite{zhang2015guided}, non-local low-rank based method (LR) \cite{li2018non}, cascaded normal regression (CNR) \cite{wang2016mesh}, deep normal filtering (DNF) \cite{li2020dnf}, and NormalF-Net (NFN) \cite{li2020normalf}.
	
	\begin{figure}[t!]
		\centering
		\includegraphics[width=\linewidth]{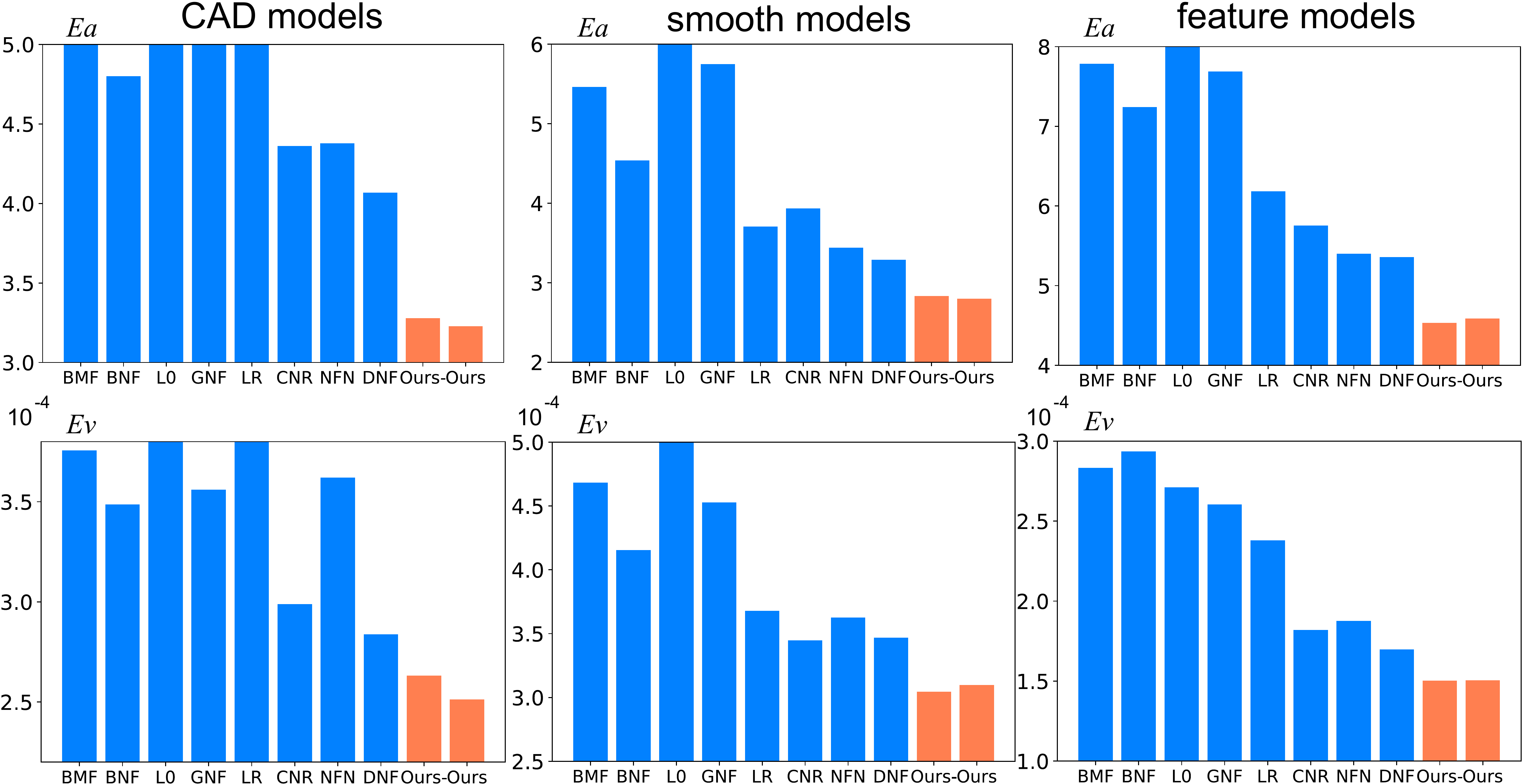}
		\caption{Quantitative comparisons with the state-of-the-art methods on the SysData benchmark dataset (i.e., CAD models, smooth models, and feature models), using the error metrics defined in Equations \ref{eq:ea} and \ref{eq:ev}. Higher bars are truncated for better illustration. ``Ours-'' denotes our method without normal refinement.}
		\label{fig:synt_quantitative}
	\end{figure}
	
	\begin{figure}[b!]
		\centering
		\includegraphics[width=\linewidth]{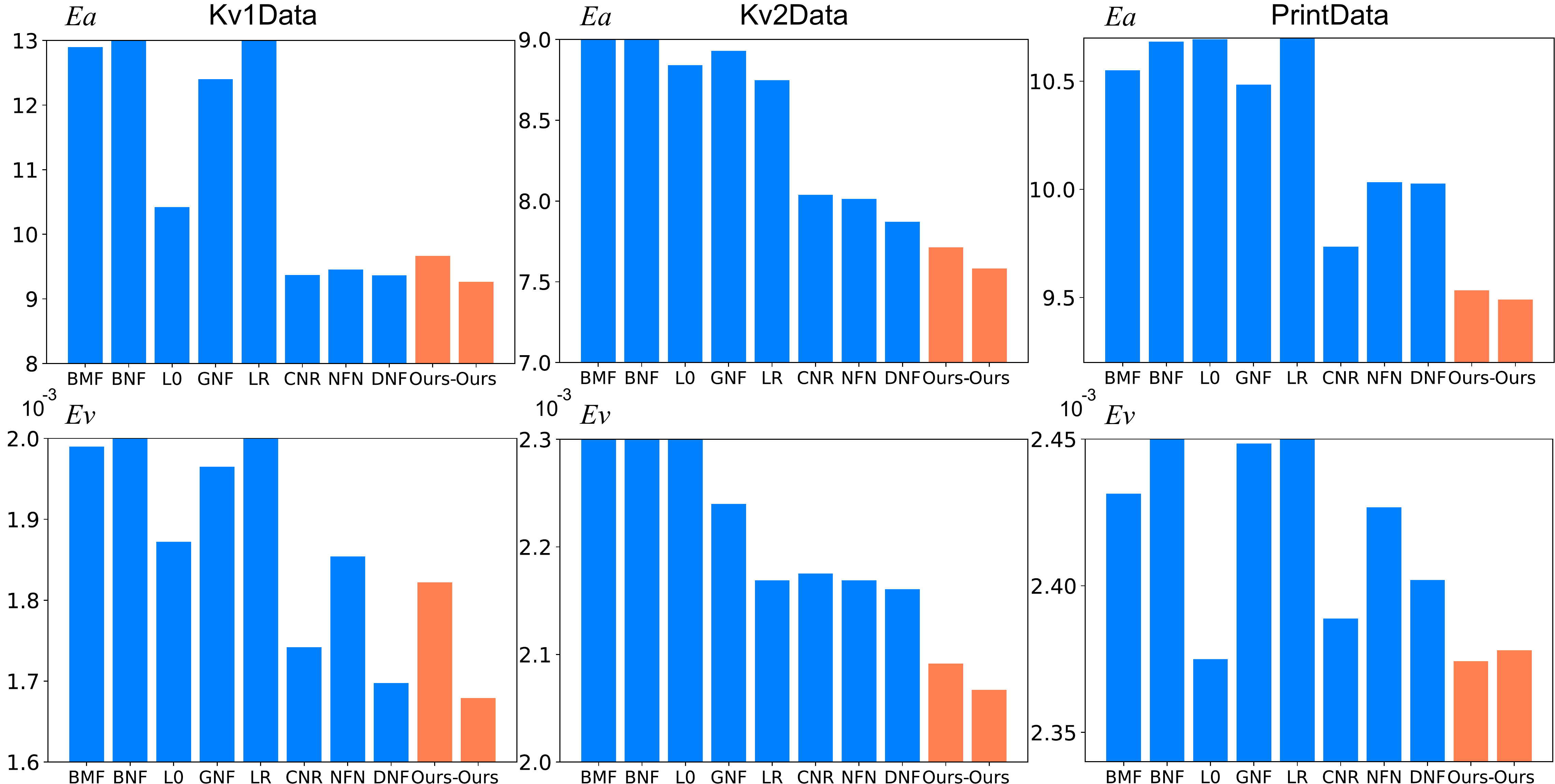}
		\caption{Comparisons with the state-of-the-art methods on the benchmark datasets: Kv1Data, Kv2Data, and PrintData. ``Ours-'' denotes our method without normal refinement.}
		\label{fig:real_quantitative}
	\end{figure}
	
	\paragraph{Synthetic Models}
	For the comparisons with the learning-based methods \cite{wang2016mesh,li2020normalf}, we use the results directly obtained from the original authors (since the authors of \cite{zhao2019normalnet} do not provide their source code, we refer to \cite{li2020normalf} for detailed comparisons therein). To compare with BMF, BNF, L0, GNF, and LR, similar to \cite{wang2016mesh}, we select the best results with fine-tuned groups of parameters for each of them as our competitors. Specifically, we use $\sigma_s=\bar{l}_e$, $\sigma_r=0.35$, $n_n = 20$, and $n_v = 20$ (20 times normal iteration and vertex updating)] in BNF, {$\lambda=0.02\bar{l_e}^2\bar{\gamma}$  ($\bar{\gamma}$ is the average dihedral angle of a mesh)} in L0, $\sigma_s=\bar{l}_e$, $\sigma_r=0.35$, $n_n = 20$, and $n_v = 20$ in GNF, $\sigma_M=0.2$, $n_n = 10$, and $n_v = 10$ in LR.
	
\begin{figure*}[h]
		\centering
		\includegraphics[width=\linewidth]{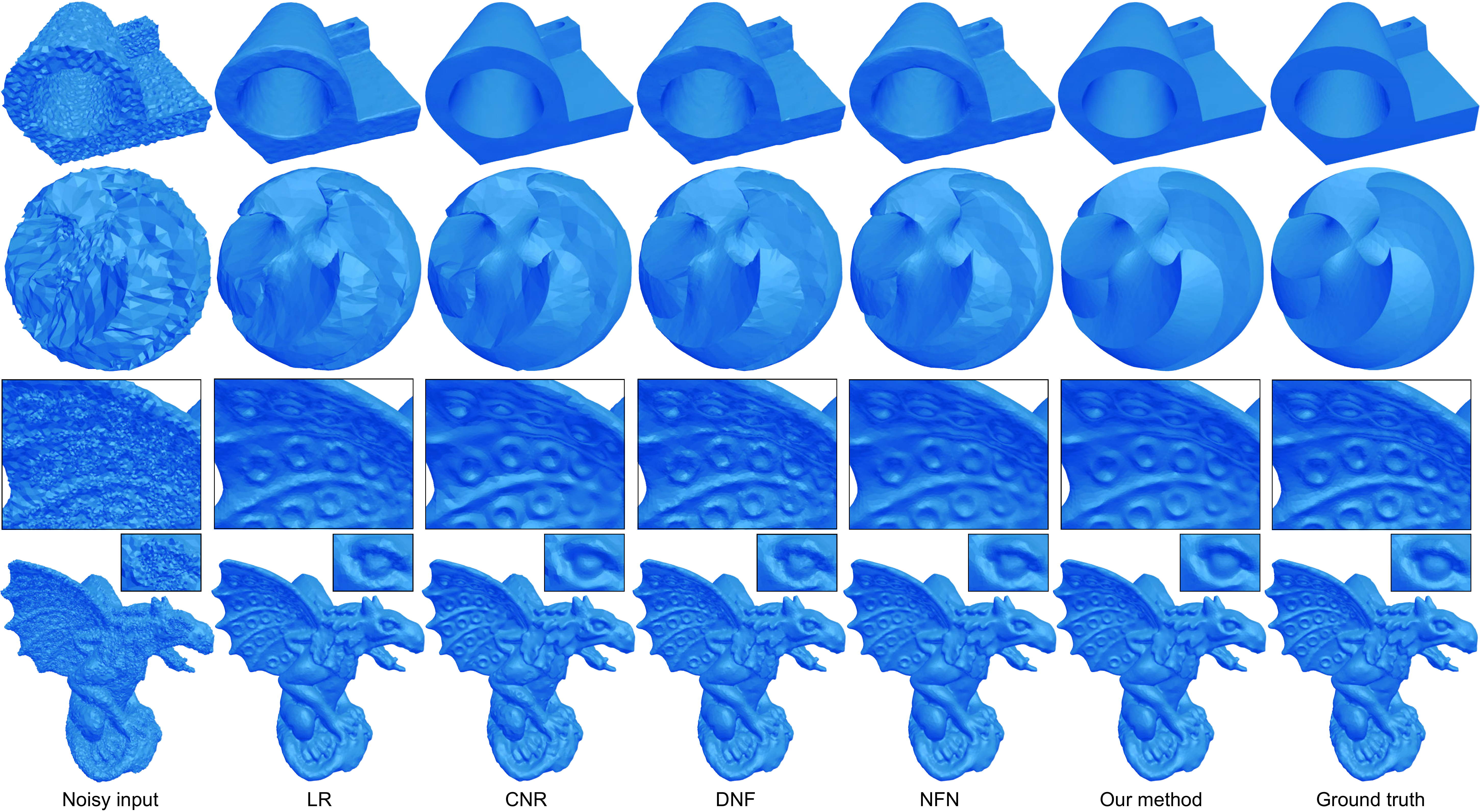}
		\caption{Visual comparisons of various methods on the SysData benchmark dataset. Models: Joint, Sharp-Shpere, Carter, and Gargoyle with the Gaussian noise of level 0.3, 0.3 and 0.3 (mean edge length),  respectively. The average normal angular errors $E_a$ (from left to right) are: (top example) $28.65^{\circ}$, $6.52^{\circ}$, $2.22^{\circ}$, $4.50^{\circ}$, $5.02^{\circ}$, and \textbf{1.86$^{\circ}$}; (middle example) $33.17^{\circ}$, $12.17^{\circ}$, $8.03^{\circ}$, $8.34^{\circ}$, $8.84^{\circ}$, and \textbf{4.39$^{\circ}$}; (bottom example) $31.78^{\circ}$, $7.56^{\circ}$, $8.72^{\circ}$, $7.98^{\circ}$, $6.89^{\circ}$, and \textbf{5.25$^{\circ}$}.}
		\label{fig:syntcompare}
	\end{figure*}

	We quantitatively compare the normal angular error $E_a$ and the vertex distance 
	error $E_v$ in the synthetic benchmark dataset SysData for all the competing methods, including ours. \figref{fig:synt_quantitative} shows the results. In this benchmark, our method consistently outperforms the compared methods in all the categories, including CAD models, smooth models, and models with rich features. Several representative visual comparisons are shown in Fig. \ref{fig:syntcompare}. In this test, the performances of NormalF-Net (NFN) \cite{li2020normalf} and the non-local recovery method (LR) \cite{li2018non} are similar because they both build upon the non-local representations. However, neither of them can handle well the CAD models with sharp features. This is a nature of the optimization-based methods, which tend to smooth out sharp features during minimization. As for  DNF (\cite{li2020dnf}), lacking mesh topological information in the normal prediction step limits its ability to remove noise. The method of cascaded regression (CNR) \cite{wang2016mesh} performs moderately well in both types of models with sharp features and fine features. Nevertheless, our method is able to recover sharp features better even from high noise, as shown in the top two rows of Fig. \ref{fig:syntcompare} while also preserving the fine-scale features as shown in the bottom two rows, clearly demonstrating the efficacy of our algorithm.
	
\begin{figure*}[h]
		\centering
		\includegraphics[width=\linewidth]{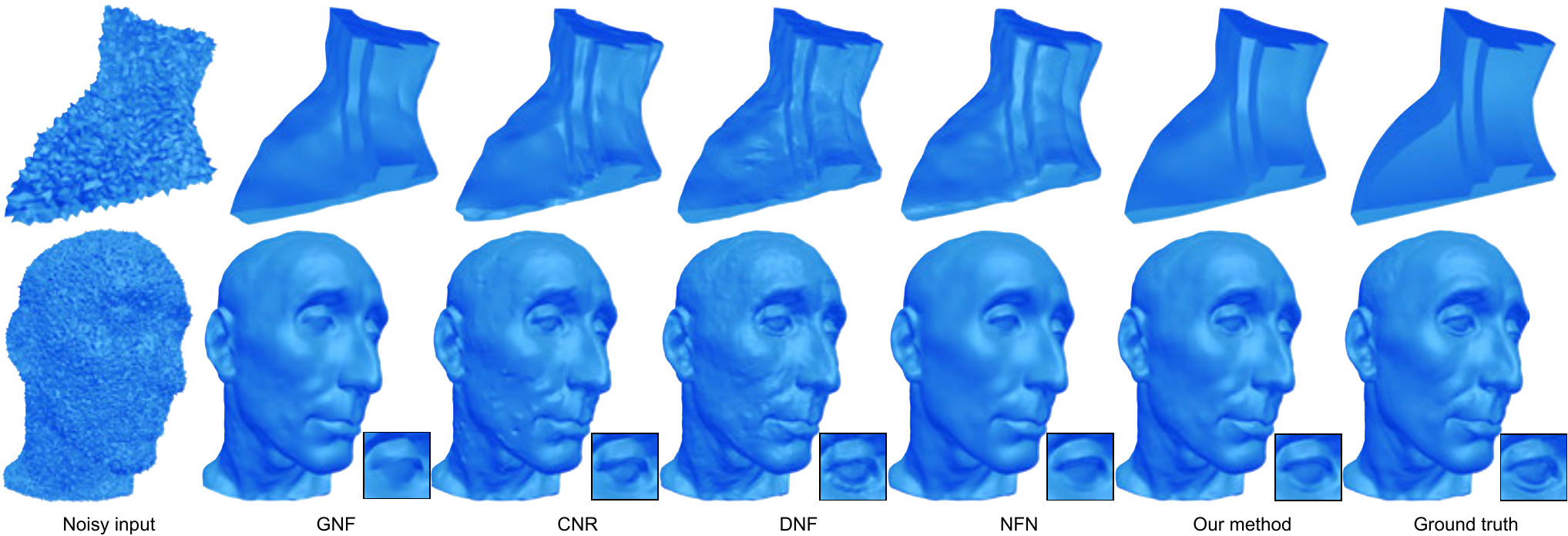}
		\caption{Visual comparisons of denoised results on models with extreme Gaussian or impulsive noise. Models: Fandisk with the Gaussian noise of level 0.6, Nicolo with the impulsive noise of level 0.6 {(both in percentage and strength)}. The average normal angular errors (from left to right) are: ($1^{st}$ row)  44.71$^{\circ}$, 6.72$^{\circ}$, 8.60$^{\circ}$, {7.52$^{\circ}$}, 11.24$^{\circ}$, and \textbf{3.87$^{\circ}$}; ($2^{nd}$ row) 36.65$^{\circ}$, 6.74$^{\circ}$, 6.99$^{\circ}$, 6.82$^{\circ}$, 5.35$^{\circ}$, and \textbf{5.03$^{\circ}$}.}
		\label{fig:morenoise}
	\end{figure*}

	Fig. \ref{fig:morenoise} shows two examples of denoised results of our method on meshes with {extremely} high-level Gaussian noise (level 0.6) and impulsive noise {(60\%, both in percentage and strength)}. The noise level in those models is significantly higher than that of any model in our training sets. Still, our method is able to denoise such noisy models and consistently performs better than the compared methods. In these examples, CNR cannot handle large positional bias between the noisy meshes and ground-truth meshes (as admitted by the authors of CNR), and thus it treats some of the impulsive and high noise as features and preserves them (e.g., see the face of the Nicolo model and the edge features on the Fandisk). In both examples, GNF \cite{zhang2015guided} tends to smooth out small features and preserve only the dominant features. On the other hand, NFN \cite{li2020normalf} performs better than CNR and GNF on the Nicolo model with fine-scale features. However, it performs worse than GNF on the Fandisk model with sharp features. This agrees with the findings above. Besides, it can be observed that DNF performs worse for high-level noises. To show the stability of our method,  we also offer an example with the irregular face resolution corrupted by two different types of noise with three different levels (first by Gaussian noise of levels 0.1 and 0.3 in different regions and then by the impulsive noise of level 0.5) in Fig. \ref{fig:normaliter}. It can be seen that our GCNs produce a high-quality denoising result, which is further improved by the normal refinement step.
	
	\begin{figure*}[h]
		\centering
		\includegraphics[width=\linewidth]{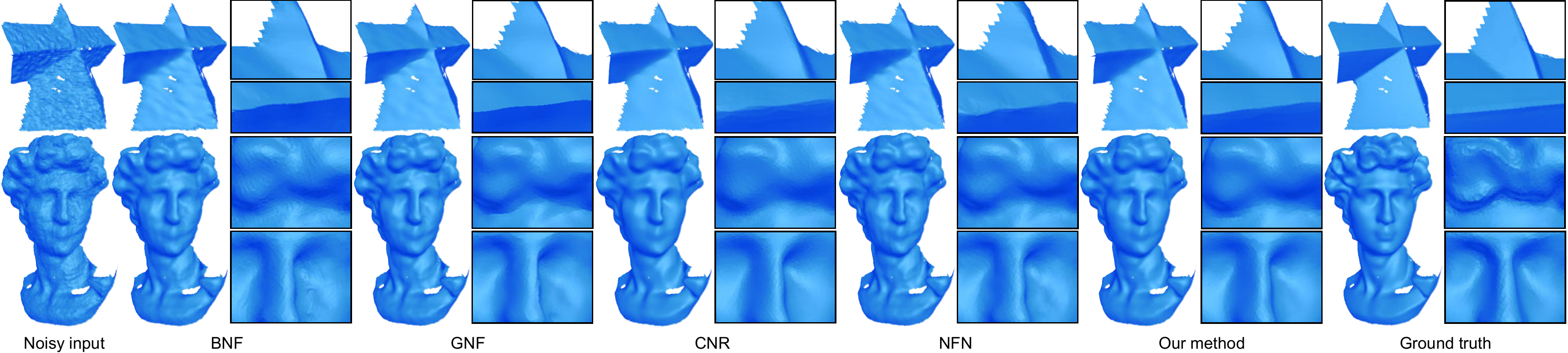}
		\caption{The denoised Kinect v2 ($1^{st}$ row) single-frame meshes and Kinect Fusion models ($2^{nd}$ row). From left to right: noisy input, denoised results of BNF \cite{zheng2010bilateral}, GNF \cite{zhang2015guided}, CNR \cite{wang2016mesh}, NFN \cite{li2020normalf}, ours and the GT. The average normal angular errors (from left to right) are: ($1^{st}$ row) $20.85^{\circ}$, $9.25^{\circ}$, $7.96^{\circ}$, $6.93^{\circ}$, $7.51^{\circ}$, and \textbf{6.58$^{\circ}$}; ($2^{nd}$ row) $17.89^{\circ}$, $12.99^{\circ}$, $11.95^{\circ}$, $11.94^{\circ}$, $12.15^{\circ}$, and \textbf{11.61$^{\circ}$}.}
		\label{fig:kinectcompare}
	\end{figure*}
	
	\begin{figure*}[h]
		\centering
		\includegraphics[width=\linewidth]{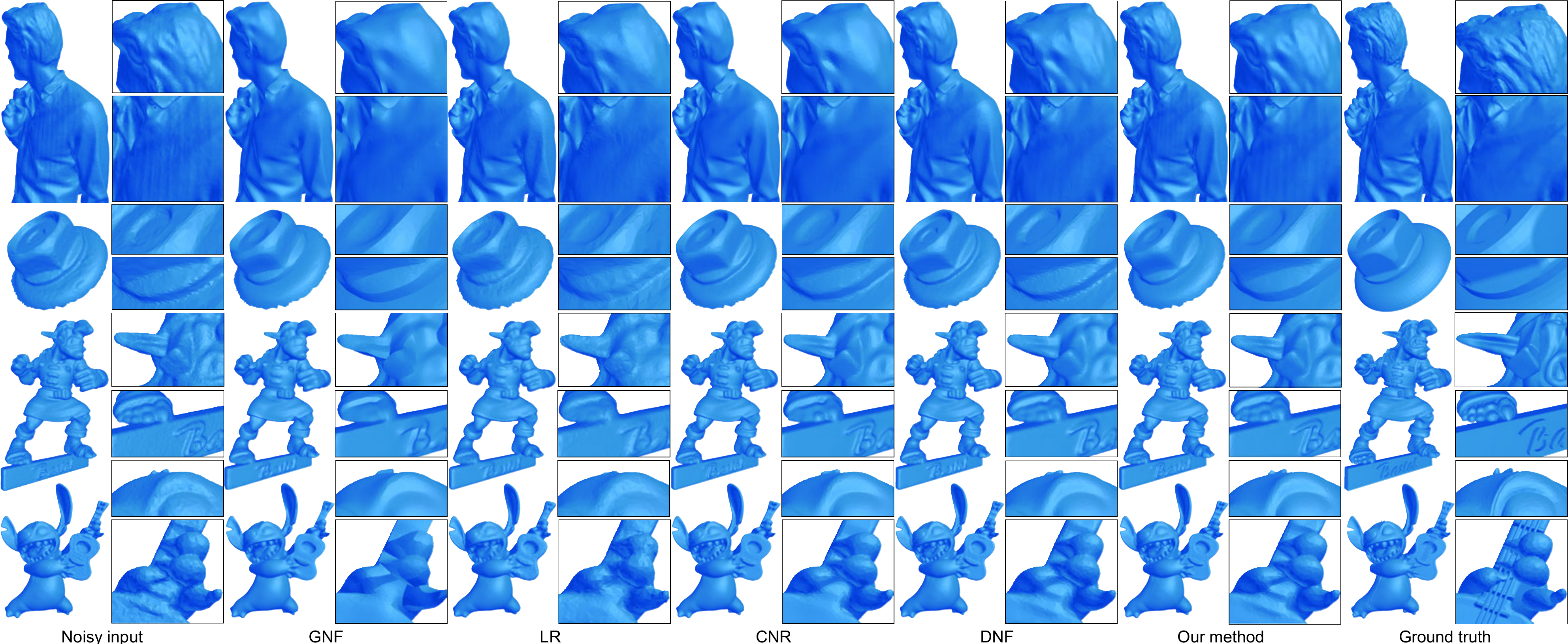}
		\caption{Denoising real-scan meshes in PrintData. From left to right: noisy input, denoised results of GNF \cite{zhang2015guided}, LR \cite{li2018non}, CNR \cite{wang2016mesh}, DNF \cite{li2020dnf}, ours, and the GT. The average normal angular errors (from left to right) are: ($1^{st}$ row) $10.88^{\circ}$, $12.06^{\circ}$, $12.27^{\circ}$, $11.02^{\circ}$, $10.74^{\circ}$, and  \textbf{10.54$^{\circ}$}; ($2^{nd}$ row) $18.08^{\circ}$, $18.68^{\circ}$, $17.96^{\circ}$, $17.23^{\circ}$, $17.67^{\circ}$, and \textbf{17.08$^{\circ}$}, ($3^{rd}$ row) $16.16^{\circ}$, $16.96^{\circ}$, $18.80^{\circ}$, $15.53^{\circ}$, $15.96^{\circ}$, and \textbf{14.95$^{\circ}$}, ($4^{th}$ row) $8.08^{\circ}$, $7.50^{\circ}$, $7.97^{\circ}$, $6.11^{\circ}$, $6.68^{\circ}$, and \textbf{5.84$^{\circ}$}.}
		\label{fig:scanres}
	\end{figure*}
	
	\paragraph{Real-scan Models} 
	For the comparisons on the real scans, we also select the best results with fine-tuned groups of parameters for BNF, $L_0$, GNF, and LR. Specifically, we set $\sigma_s=\bar{l}_e$, $\sigma_r=0.45$, $n_n = 20$, and $n_v = 20$ in BNF, $\lambda=0.06\bar{l_e}^2\bar{\gamma}$  in $L_0$, $\sigma_s=\bar{l}_e$, $\sigma_r=0.45$, $n_n = 20$, and $n_v = 20$ in GNF, and $\sigma_M=0.4$, $n_n=10$, and $n_v=10$ in LR. 
	
	The quantitative comparisons are shown in Fig. \ref{fig:real_quantitative}. Our method also consistently achieves the best results in all Kv1Data, Kv2Data, and PrintData real-scan benchmarks. The representative results of qualitative comparisons on Kv2Data and K-FData are shown in Fig. \ref{fig:kinectcompare}. Note that in these real scans, due to the low capture quality, the fine features of the acquired models are often highly corrupted or even destroyed by noise (e.g., the eyes and the mouth in the statue of Fig. \ref{fig:kinectcompare} have been somehow erased and are difficult to perceive even for humans). Thus it is difficult for most of the denoising methods to faithfully recover these features, so is ours. Nonetheless, our method still receives the lowest reconstruction errors while also preserving the features better than the previous methods. Moreover, the models in Kv1Data, Kv2Data, and K-FData often do not exhibit complex geometric features and are not very suitable for testing the limit of the feature-preserving ability of different methods.
	
	To further demonstrate the performance of our method, we apply denoising on the proposed PrintData. As shown in Fig. \ref{fig:scanres}, the scans in our dataset often involve more geometric details than the scans by Kinect but still suffer from noise of small scales. The Kv1Data, Kv2Data, and K-FData use high-resolution scanned models (by an Artec $\text{Spider}^{\text{TM}}$ scanner) as ground truth and low-precision Kinect scanned models as noisy input. In contrast, we take existing 3D digital models as our ground truth and take the scanned models as noisy input. Due to the quality of 3D printing (it may introduce larger errors than 3D scanning), there are naturally small differences between the printed models and the ground truth, e.g., the vertical lines in the first row of Fig. \ref{fig:scanres} (the corresponding printed result is shown in Fig. \ref{fig:printdataset}). Our results preserve the original details and meanwhile have the lower errors. Also, our results have better performance around the sharp features especially in CAD models (the second row in Fig.\ref{fig:scanres}). Again, our approach consistently preserves the features (e.g., in the ear of the Goblin and the bottom letters shown in the third row of Fig. \ref{fig:scanres} and the head of the Stitch and the guitar in the fourth frow of Fig. \ref{fig:scanres}) better than the compared methods.
	
	\begin{table}
		\renewcommand{\arraystretch}{1.3}
		\caption{Time comparisons with the state-of-the-art learning-based denoising methods: NormalNet (NN) \cite{zhao2019normalnet}, CNR \cite{wang2016mesh}, DNF \cite{li2020dnf}, and NormalF-Net (NFN) \cite{li2020normalf}. The running time is in seconds. Please find the corresponding models in the $2^{nd}$ in \figref{fig:syntcompare}, the rightmost one in \figref{fig:dataset} (f), the bottom one in \figref{fig:dataset} (h), and the $4^{th}$ in \figref{fig:syntcompare}.}
		\centering
		\begin{tabular}{c|c|c|c|c}
			\hline
			\bfseries Model & Sharp-Sphere & Fertility & Eros & Gragoyle\\
			\hline
			\bfseries Faces & 20882 & 27954 & 100000 & 171112\\
			\hline
			\hline
			\bfseries NN & 836.12s & 1132.42s  & 9163.24s & 20763.28s\\
			\hline
			\bfseries CNR & 1.27s & 1.71s & 5.16s & 10.04s\\
			\hline
			\bfseries DNF & 352.39s & 494.94s & 1912.74s & 3376.69s\\
			\hline
			\bfseries NFN & 97.37s & 109.63s & 480.55s & 975.05s\\
			\hline
			\bfseries Ours & 11.88s & 12.97s & 59.95s & 145.96s\\
			\hline
		\end{tabular}
		\label{tbl:runtime}
	\end{table}
	
	\paragraph{Timing}
	Following \cite{wang2016mesh}, we train our network for synthetic and real scans separately. Our method takes about 6-8 hours for training the network of synthetic models and 2-3 hours for training the network of real scan models.
	
	Our learning-based method is also efficient at runtime and its performance comparisons with the previous learning-based methods on typical models are summarized in Table \ref{tbl:runtime}. The runtime largely depends on the complexity of network architectures. It can be seen that our method runs one magnitude faster than DNF \cite{li2020dnf} and NormalF-Net \cite{li2020normalf} and two magnitudes faster than NormalNet \cite{zhao2019normalnet}, which is based on 3D convolutions on voxels. The MLP-based method \cite{wang2016mesh} receives the best performance due to its simplicity in the network architecture (with only a few fully connected layers). Nonetheless, our method achieves the best denoising quality while being well balanced between the efficacy and efficiency among the learning-based methods.
	
	\subsection{Ablation Studies}\label{subsec:ablation}
	In this subsection, we introduce the ablation studies to show the impact of various algorithmic components of our method.
	
	\paragraph{Patch Size}
	The size of patches {$r$ in \equref{eqn:patchsize}} determines the receptive filed of our GCNs, i.e., how much local geometry we can see in a patch w.r.t. an entire mesh. This is an important hyper-parameter. Essentially, $r$ should be set to a fixed value, e.g., $5\%$ of the diagonal length of the mesh's bounding box to make the patch sizes consistent. In our implementation, we set $r=ka_i^\frac{1}{2}$ as a value relevant to the triangle area to accommodate different surface samplings. Besides, the input to our GCN should be padded into a fixed size of nodes since we train the GCN with batches.
	
\begin{figure}[t!]
		\centering
		\includegraphics[width=\linewidth]{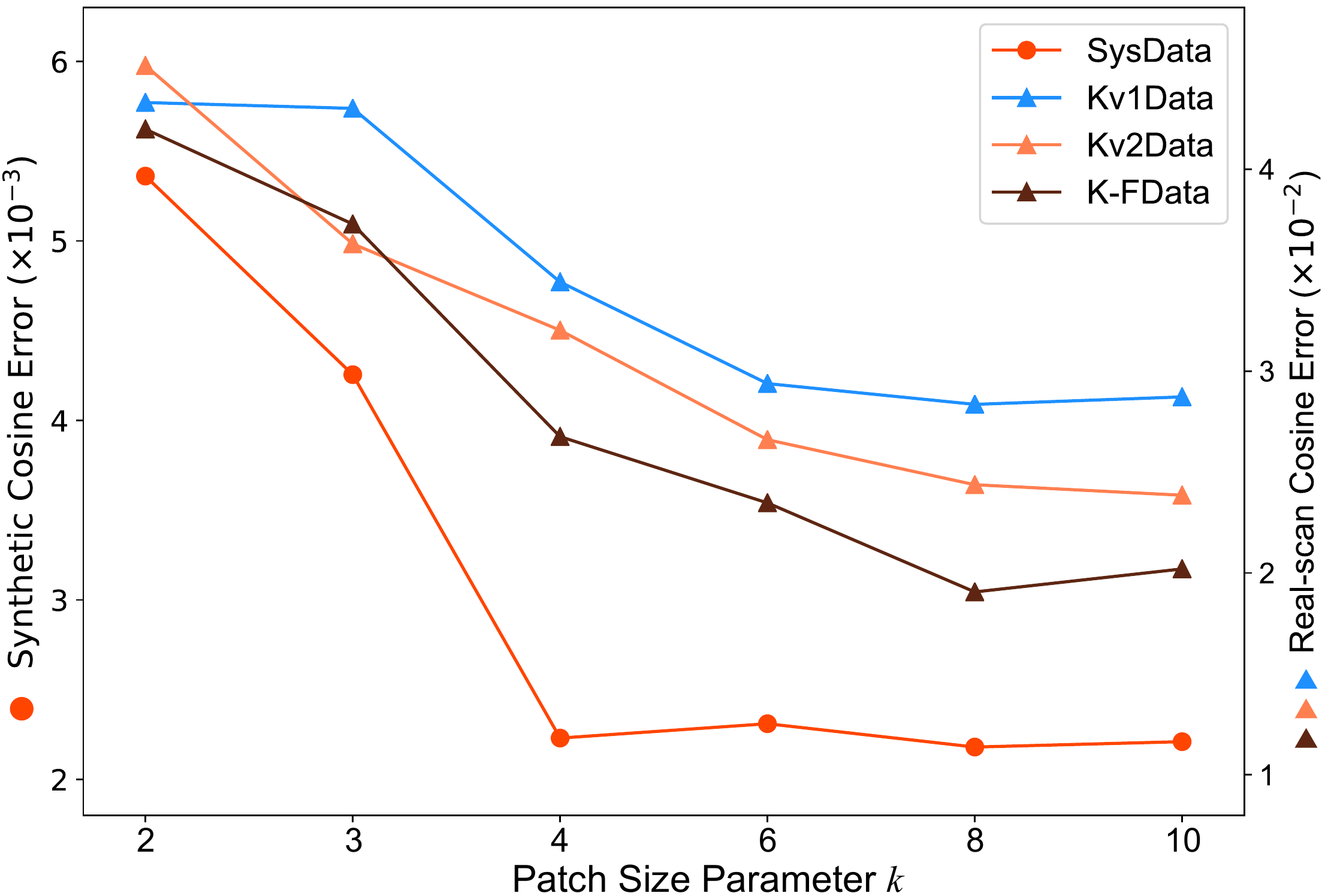}
		\caption{Impact of patch size parameter $k$ on different datasets. For the synthetic data (SysData), a value of 4 leads to satisfactory results while for the real scan data (i.e., Kv1Data, Kv2Data, K-FData), $k=8$ leads to sufficient good results.}
		\label{fig:patchsize}
	\end{figure}

\begin{figure}[t!]
		\centering
		\includegraphics[width=\linewidth]{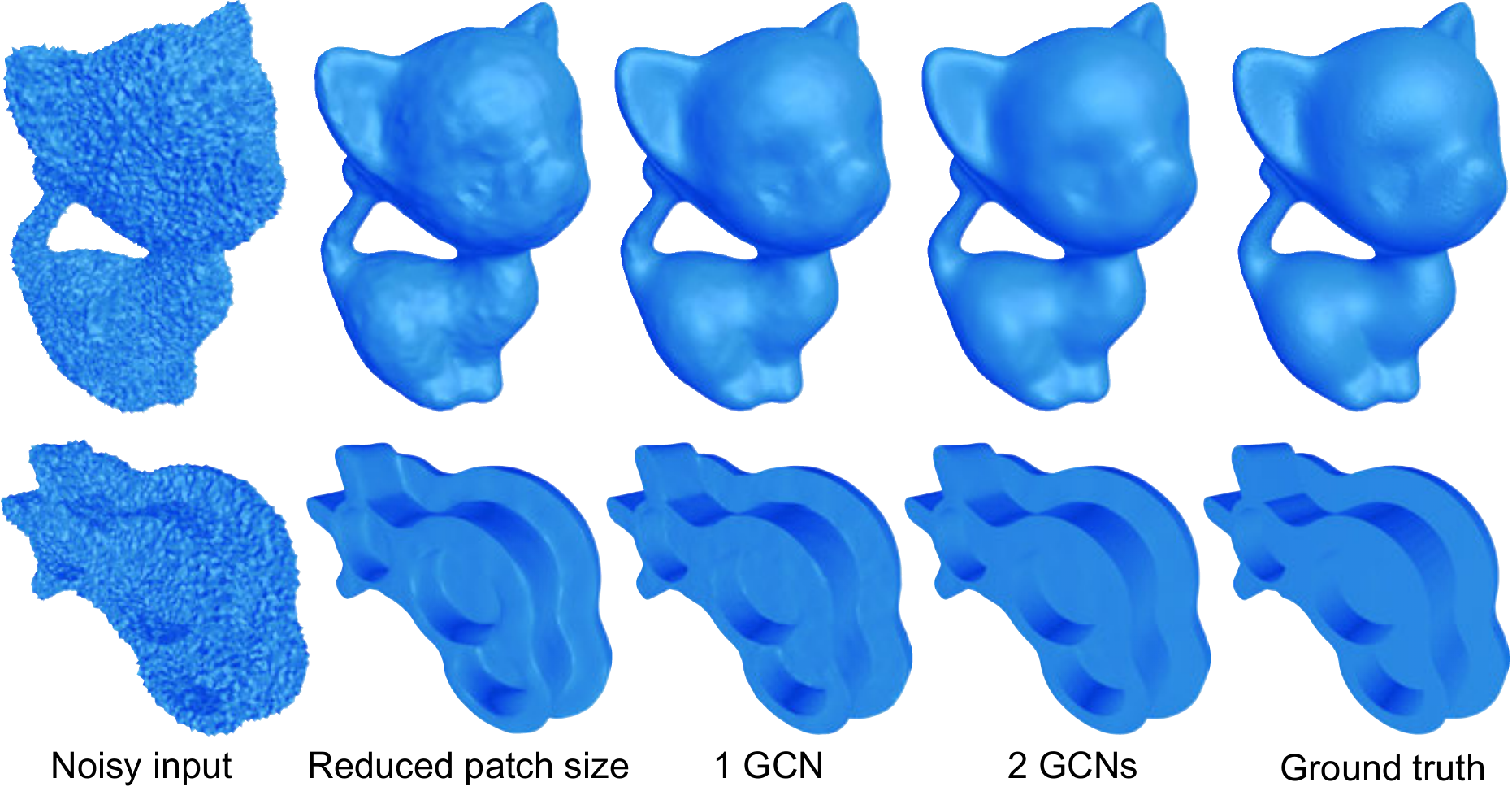}
		\caption{Comparative denoised results of different patch sizes ($2^{nd}$ column: $k=2$) and different numbers of cascaded GCNs. The average angular errors (from left to right) are: ($1^{st}$ row) 40.52$^{\circ}$, 5.58$^{\circ}$, 3.84$^{\circ}$, and \textbf{3.29$^{\circ}$}; ($2^{nd}$ row) 33.41$^{\circ}$, 5.61$^{\circ}$, 2.47$^{\circ}$, and \textbf{2.01$^{\circ}$}.}
		\label{fig:ablatioinvis}
	\end{figure}

	We test over six scales of patches, i.e., $k=2,3,4,6,8$, and $10$ ($k$ is defined in \secref{subsec:select}). For training in batches, we fix the number of graph nodes to 16, 32, 64, 128, 256, and 512 according to the values of $k$ (in case the number of facets within the patch does not equal to the defined number, we perform random shrinkage or extension to the local patch). Fig. \ref{fig:patchsize} shows the corresponding performance of these choices. We find that setting $k=4$ makes our GCN regress well enough for the synthetic data, while setting $k=8$ is sufficient for the real-scan Kinect data. A desired value of $k$ for the real-scan Kinect data is larger, mainly because the surface resolution of the real-scan data is often much larger than that of the synthetic ones and the low-precision scans often bring in large-scale noise. The experiments show that as long as the patch sufficiently covers local regions, the performance is stable. A visualization effect of meshes denoised with different patch sizes is shown in \figref{fig:ablatioinvis}.
	
\begin{figure}[t!]
		\centering
		\includegraphics[width=\linewidth]{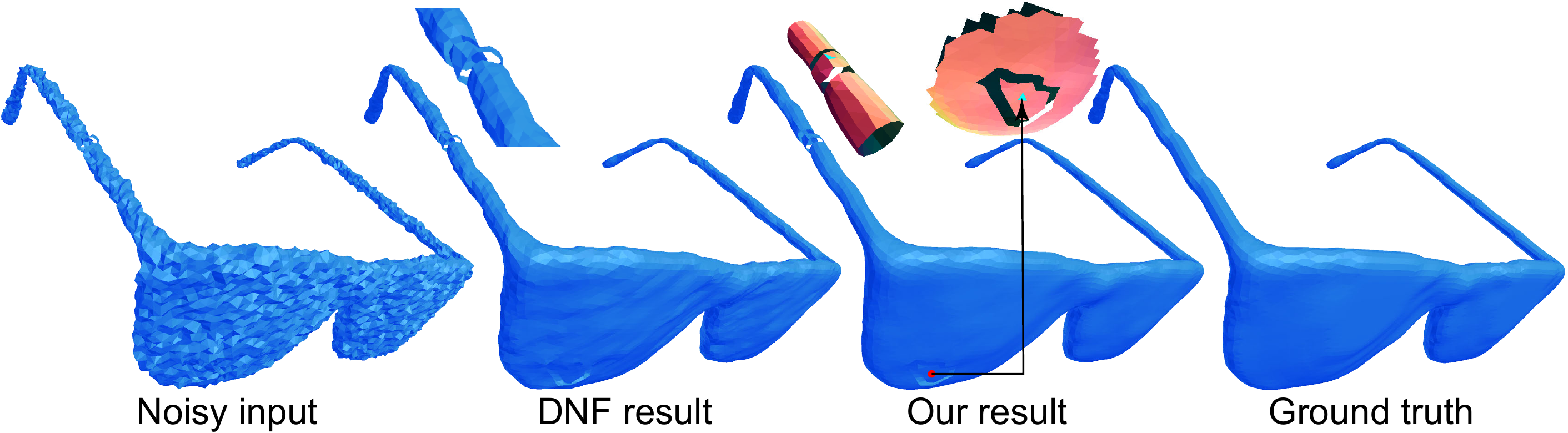}
		\caption{A denoising result of a broken thin-structured model. Our method is able to learn features well even from patches with disconnected graph structures. The average normal angular errors $E_a$ are: $28.114^{\circ}$ (input), $7.35^{\circ}$ (DNF's result), and \textbf{4.70$^{\circ}$} (our result).}
		\label{fig:thin_structure}
	\end{figure}

	Since our patches are generated within a sphere, some noisy, disconnected facets might be included in our patch graphs if the geometry around a facet has a broken or thin structure. Fig. \ref{fig:thin_structure} shows such an example, where facets in the disconnect regions and on the other side of the lens are presented in the same patch graph. Here $k=8$ is used. Nevertheless, unlike DNF \cite{li2020dnf}, our method is not sensitive to such thin or broken structures and produces satisfactory results, as shown in Fig. \ref{fig:thin_structure}. This is due to that our static EdgeConv inherently exploits a mesh's original structure and our dynamic EdgeConv distinguishes which faces features are helpful, thus making our method robust to such noisy representations.
	
	\begin{figure}[h]
		\centering
		\includegraphics[width=\linewidth]{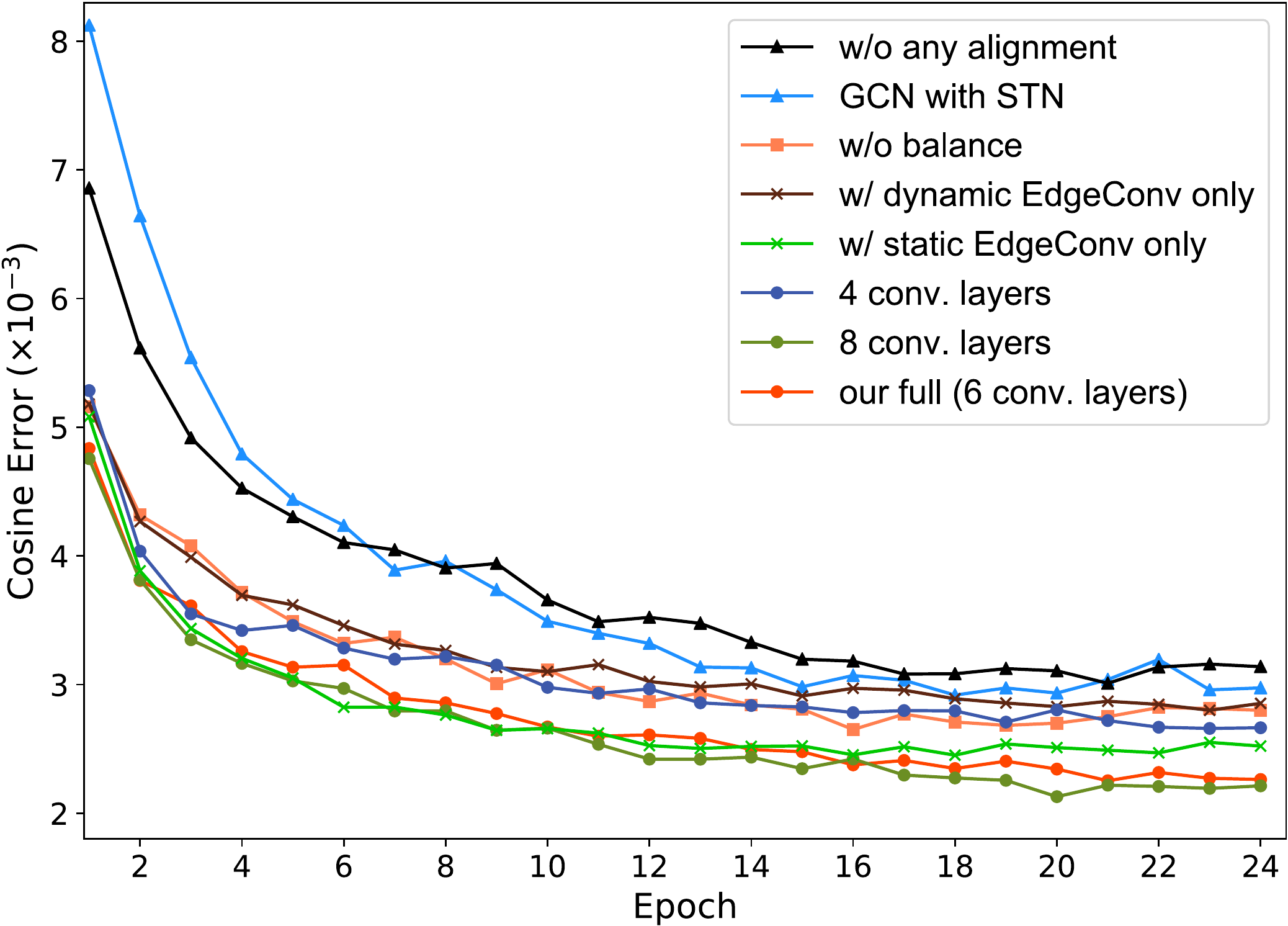}
		\caption{Ablation studies on various design choices of our algorithm.}
		\label{fig:loss}
	\end{figure}
	
	\begin{figure}[h]
		\centering
		\includegraphics[width=\linewidth]{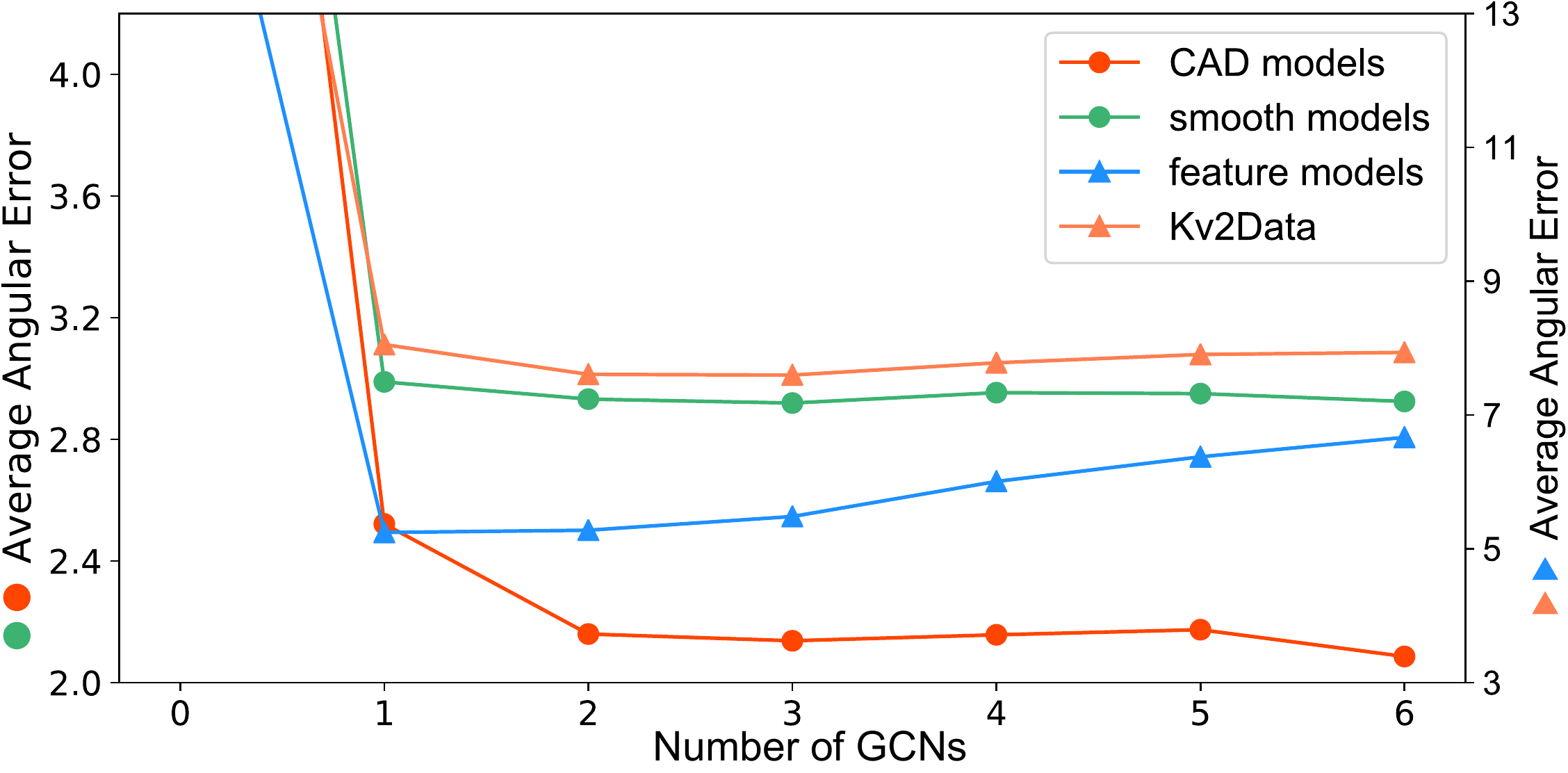}
		\caption{Test on the number of GCNs. Cascading a larger number of GCNs does not necessarily increase the performance. In fact, it might over-smooth certain fine-scale features when this number increases.}
		\label{fig:cascade}
	\end{figure}
	
	\paragraph{Number of Graph Convolution Layers}
	To examine how the number of layers (i.e., how deep) of our GCN affects the results, we conduct the following experiments. For the number of layers of graph convolution, i.e., $L_e + L_d$, we test the following numbers: 4 ($L_e, L_d=2$), 6 ($L_e, L_d=3$), and 8 ($L_e, L_d=4$). We set $L_e=L_d$ for combinatorial simplicity. Fig. \ref{fig:loss} plots the performance. We find that the performance does not increase much after 6, showing that a moderate size of GCN is sufficient in learning the local geometry details of our aligned patches. Thus we use {$L_e = 3$ and $L_d = 3$} for our first GCN and {$L_e = 2$ and $L_d = 2$} for the rest.
	
	\paragraph{Number of GCNs}
	In this test, we show the necessity of multiple GCNs. To do so, we randomly sample a set of representative models in each category of the SysData {benchmark dataset}, including 4 CAD models, 4 smooth models, and 4 models with rich features {(0.1--0.3 levels of Gaussian noise are added to each model)}. We also sample a similar set of models in the {Kv2Data benchmark dataset}. We then run the test on these sampled sets using 1 -- 5 GCNs for denoising. The quantitative comparisons of the performance are shown in Fig. \ref{fig:cascade}. It can be seen that the performance of adding more GCNs stops improving when the number is greater than 2 for the CAD models, smooth models, and the Kinect v2 models with low-frequency features. For the models with rich fine features, adding more GCNs might result in an over-smoothed effect since under such circumstances, it is hard to distinguish those fine features from noise. Hence, we use 2 GCNs for denoising in all our experiments. Visual results of using 1 and 2 GCNs are shown in \figref{fig:ablatioinvis}.
	
	\paragraph{Patch Alignment} To show the influence of our patch alignment, we replace our tensor voting with a spatial transform network (STN) \cite{wang2019dynamic} module implemented in our GCN. STN is widely used in point cloud processing works \cite{qi2017pointnet, wang2019dynamic} to eliminate spatial variations among the inputs. Fig. \ref{fig:loss} {(the curve of ``GCN with STN'')} shows that STN does not work as well as our normal tensor voting. This agrees with the finding {that adding a spatial transformation module does not improve the performance much} in \cite{qi2017pointnet} and partially proves that spatial transformation is indeed not easy to learn by neural networks. On the other hand, if we do not use any alignment scheme, the performance decreases (see the curve of ``w/o any alignment'' in Fig. \ref{fig:loss}). Patch alignment may be  influenced by the noise level, however, our cascaded optimization may help progressively correct the errors (\figref{fig:ablatioinvis}).
	
	\paragraph{Data Balancing} We also examine the influence of our data balancing strategy. Fig. \ref{fig:loss} shows that the performance of our network increases slightly due to the adopted data balancing strategy. Let us denote $r$ as the ratio of the number of featured facets compared to that of non-feature ones. Without data balancing, the value of $r$ is approximately $0.1$. We test over $r=\{0.5,1,1.5,2,5,10\}$ and find that the performance increases very slightly after $k=1.5$ and starts to decrease after $r=10$. Hence, throughout our experiments, we use $r=1.5$ for data balancing. The data balancing allows more effective learning of the underlying features.
	
	\paragraph{Static and Dynamic Graph Convolutions}
	We train our GCNs with both static and dynamic EdgeConv. To prove their effectiveness, we examine the following alternatives: a network with static EdgeConv only and a network with dynamic EdgeConv only. Again, it is witnessed that a combination of the two allows more information flow from both neighboring graph nodes and these potentially unconnected ones, thus leading to a more effective learning of the features, as shown in Fig. \ref{fig:loss}. On the other hand, the network with only EdgeConv leads to a better performance than the one with only dynamic EdgeConv. This shows that the original graph structure in the mesh is already very informative.
	
	\paragraph{KNN}
	We use KNN to dynamically construct graph structures in dynamic EdgeConv. We examine the influence of the performance on different values of $K$. We test $K=4,8,12$, and $16$. The performance of $K=4$ is similar to that of using static EdgeConv only and does not improve after $K=8$. Thus we use $K=8$ throughout our experiments.
	
	\subsection{Implementation Details}
	As mentioned before, our GCNs have different numbers of convolution and MLP layers. In our experiments, the numbers of feature channels are set as (64, 128, 128, 256, 256, 256, 1024, 512, 256, 64) in the first GCN and (64, 128, 256, 256, 512, 256, 64) in the other GCNs. In the training stage,  we use Adam ($\beta_1=0.9$ and $\beta_2=0.999$) for optimization with the base learning rate 0.0001. We set the batch size as 128 and train 24 epochs for the first GCN, and 16 epochs for the rest. At runtime, we also regress normals in batches. Due to the limited GPU memory, we set the batch size as 720 for patch size $k=4$ and set the batch size as 160 for patch size $k=8$. 
	
	\section{Limitations}
	Our method has several limitations. First, since we learn the unknown noise patterns from massive data, the capacity of our method is limited by the training data. Second, although our method is able to generalize to unseen noise levels, it could still fail to recover the underlying features once they are deeply corrupted by noise. Such effects have been demonstrated in examples of this paper with extremely high noise {(Fig. \ref{fig:morenoise})} or low quality noisy input from low-end depth cameras {(Fig. \ref{fig:kinectcompare}).} Third, one assumption of our method is that the geometry variation of the noise and that of the underlying features are different so that both variations can be well modeled by our GCNs. If this assumption is broken, our method would fail to distinguish features from noise and tend to either smooth out the features or preserve the noise. This is often the case with the meshes containing many fine-scale features (Fig. \ref{fig:teaser}). A more robust and fine-grained classification module might be helpful but cannot completely solve this ill-posed problem due to the inherent ambiguity. Fourth, like most of the existing feature-preserving denoising methods, our method does not change the mesh connectivity, thus we cannot remove topological noise. Utilizing dynamic graph convolution that explicitly updates the mesh connectivity may be helpful to resolve this issue. We consider it as an orthogonal future work.
	
	\section{Conclusion and Discussion}\label{sec:con}
	In this paper, we have presented the first GCN-based approach for feature-preserving mesh denoising. Our method takes a triangular mesh as input and employs multiple GCNs to progressively regress the noise-free normals of the underlying surface patches. An essential ingredient of our method is to represent the local surface patches as graphs in the dual space of triangles. We show such an intact representation allows convolution operations to be performed directly on the mesh surface to effectively learn geometric features. We employ both static and dynamic graph convolutions to aggregate features from both connected neighbors and unconnected ones, enabling a more effective feature learning. Extensive experimental results show that our GCN models achieve the new state-of-the-art results while being well balanced between efficacy and efficiency. 
	
	Although our current implementation relies on triangular meshes, it can be easily adapted to other representations, for instance, quad meshes. We are also interested in extending our method for denoising unorganized point clouds or non-manifold meshes. For point clouds, it might be tricky if we directly perform denoising on normals since the subsequent vertex updating step is not feasible if the connectivity is unknown. One possibility is to regress the point positions with dynamic EdgeConv \cite{rakotosaona2020pointcleannet}, but in a local and progressive manner. For non-manifold meshes, we may require additional such training data and a new vertex updating scheme. Moreover, we believe that our framework can be extended for  applications such as geometric texture synthesis, mesh feature enhancement, mesh topological noise removal, and shape deformation. We also believe that the general framework of regressing a complex function over a 3D surface in a cascaded and local manner could be inspiring for various geometry tasks, such as surface reconstruction \cite{jiang2020local} and super-resolution.
	
	
	\begin{acks}
		We would like to thank the anonymous reviewers for their constructive comments. This work was supported in part by the National Key Research \& Development Program of China (2018YFE0100900) and the NSF China (No. 61890954, 61772024, 61732016). The work of Evgeny Burnaev in Sections \ref{sec:patch}-\ref{sec:exp} related to learning of neural networks was supported by Ministry of Science and Higher Education (No. 075-10-2021-068).
	\end{acks}
	
	\bibliographystyle{ACM-Reference-Format}
	\bibliography{cited}
	
	\appendix
	
	
	
\end{document}